\documentclass[preprint,prd,aps,showpacs,showkeys,nofootinbib]{revtex4}
\usepackage{graphicx}
\usepackage{dcolumn}
\usepackage{bm}
\begin{document}
\title{The corrections from one loop and  two-loop Barr-Zee type \\diagrams to muon
MDM in BLMSSM}
\author{Shu-Min Zhao$^1$\footnote{zhaosm@hbu.edu.cn}, Tai-Fu Feng$^{1}$\footnote{fengtf@hbu.edu.cn}, Hai-Bin Zhang$^{1,2}$, Ben Yan$^{1}$, Xi-Jie Zhan$^1$}
\affiliation{$^1$ Department of Physics, Hebei University, Baoding 071002,
China\\
$^2$ Department of Physics, Dalian University of Technology, Dalian 116024, China}
\date{\today}
\begin{abstract}
In a supersymmetric extension
of the standard model where baryon
and lepton numbers are local gauge symmetries(BLMSSM) and the Yukawa couplings between Higgs doublets and exotic quarks are considered,
 we study the one loop diagrams and the two-loop Barr-Zee type diagrams with a closed Fermi(scalar)
 loop between the vector Boson and Higgs. Using the effective Lagrangian method,
we deduce the Wilson coefficients of dimension 6 operators contributing to the anomalous magnetic
moment of muon, which satisfies the electromagnetic gauge invariance. In the numerical analysis,
we consider the experiment constraints from Higgs and neutrino data. In some parameter space, the new physics
contribution is large and even reaches $24\times10^{-10}$, which can remedy the deviation well.

\end{abstract}


\maketitle

\section{Introduction}
\indent\indent
 The magnetic dipole moment (MDM) of lepton has close relation
 with the new physics beyond the standard model(SM). The current world average value \cite{pdg2012} of $(g-2)_{\mu}$ experiment is
 \begin{eqnarray}
 a^{exp}_{\mu} = \frac{1}{2}
(g_{\mu}-2) = 11659208.9(5.4)(3.3)\times10^{-10}.
\end{eqnarray}
There are three type contributions to the MDM of
 muon \cite{3type} such as: QED loops, hadronic contributions and electroweak corrections. The SM theoretical prediction of muon MDM is\cite{smtheory}
\begin{eqnarray}
a^{SM}_{\mu}= 11659184.1(4.8)\times10^{-10}.
\end{eqnarray}
The deviation between the SM prediction and experimental result is given as the follows, which  lies in the range of
$\sim3\sigma$\cite{wucha}.
\begin{eqnarray}
\Delta a_{\mu} =a^{exp}_{\mu}-a^{SM}_{\mu}= 24.8(8.7)(4.8)\times 10^{-10}.
\end{eqnarray}

The minimal supersymmetric extension of the standard model
(MSSM) \cite{MSSM} is one of the most attractive candidates in the models beyond the SM, and draws
physicists most attentions for a long time.
 A minimal supersymmetric extension of
the SM with local gauged B and L(BLMSSM) is a favorite one, because it has two advantages.
1. The broken baryon number (B) can explain asymmetry of matter-antimatter in the universe.
2. The neutrinos should have tiny mass from the neutrino oscillation experiment.
 In theory, the tiny mass can be induced from the heavy majorana neutrinos by the seesaw mechanism.
Therefore, at some scale the lepton number (L) should be broken too.

Extending SM, with B and L as
spontaneously broken gauge symmetries around ${\rm TeV}$ scale the models are studied\cite{BLMSSM}.
Neglecting the Yukawa couplings between Higgs doublets and exotic quarks
in BLMSSM, the authors study the lightest CP-even Higgs \cite{BLMSSM,BLMSSM1}.
In the BLMSSM, considering the Yukawa couplings between Higgs and exotic quarks,
we study the lightest CP-even Higgs$(h^0)$ mass and the decays  $h^0\rightarrow\gamma\gamma$, $h^0\rightarrow ZZ (WW)$\cite{weBLMSSM}, which are also studied in other models. In the CP-violating BLMSSM, the neutron electric dipole moment(EDM) is investigated\cite{smneutron}.

To find new physics beyond the SM,
research the MDMs \cite{MDM,MDM1} and  EDMs\cite{EDM} of leptons are the effective ways.
There are some works for the supersymmetric (SUSY) one-loop contributions to muon MDM, and in some parameter space\cite{one-loopsusy}
 the numerical results can be large. In $\mu\nu MSSM$, we study the muon MDM at one-loop level\cite{wucha}.
 The authors investigate two-loop Barr-Zee-type diagrams\cite{barrzee} and obtain the electric
dipole moments (EDMs) and MDMs of light fermions.  Using the heavy mass expansion approximation (HME) and the projection operator method, the authors show two-loop standard
electroweak corrections to muon MDM \cite{heavymass}. There are also several works about the muon MDM from two-loop diagrams \cite{Heinemeyer,feng} in SUSY model.

In this work, we study the one loop diagrams and two-loop Barr-Zee type diagrams with a closed  scalar (Fermi)
 loop between vector Boson and Higgs in the frame work of BLMSSM. Taking into account the Yukawa couplings between Higgs doublets and exotic quarks,
 we investigate these contributions to muon MDM with the effective Lagrangian method.
Using the same method as in the Ref.\cite{feng}, we deduce all dimension 6 operators
and their coefficients.
Attaching a photon in all possible ways on the internal line of one self-energy diagram,
one can obtain the corresponding triangle diagrams, and the sum of these amplitudes satisfies the Ward identity
required by the QED gauge symmetry.
Adopting the equations of
motion to external leptons, we can neglect higher dimensional operators(dimension 8
operators) safely.

 After this introduction, we briefly summarize the main ingredients
of the BLMSSM, and show the needed couplings for exotic leptons and exotic quarks in section 2.
 We collect the one-loop and two-loop corrections to the muon MDM
in section 3.
Section 4 is devoted to the numerical analysis and
discussion for the dependence of muon MDM on
the BLMSSM parameters. In section 5, we give our conclusion. Some
formulae are collected in the appendix.

\section{Some coupling in BLMSSM}
Physicists enlarge the SM with the local gauge group of
$SU(3)_{C}\otimes SU(2)_{L}\otimes U(1)_{Y}\otimes U(1)_{B}\otimes U(1)_{L}$, and obtain BLMSSM \cite{BLMSSM}.
To to cancel $L$ and $B$ anomaly, the exotic leptons ($\hat{L}_{4}\sim(1,\;2,\;-1/2,\;0,\;L_{4})$,
$\hat{E}_{4}^c\sim(1,\;1,\;1,\;0,\;-L_{4})$, $\hat{N}_{4}^c\sim(1,\;1,\;0,\;0,\;-L_{4})$,
$\hat{L}_{5}^c\sim(1,\;2,\;1/2,\;0,\;-(3+L_{4}))$, $\hat{E}_{5}\sim(1,\;1,\;-1,\;0,\;3+L_{4})$,
$\hat{N}_{5}\sim(1,\;1,\;0,\;0,\;3+L_{4})$) and
 the exotic quarks ($\hat{Q}_{4}\sim(3,\;2,\;1/6,\;B_{4},\;0)$,
$\hat{U}_{4}^c\sim(\bar{3},\;1,\;-2/3,\;-B_{4},\;0)$,
$\hat{D}_{4}^c\sim(\bar{3},\;1,\;1/3,\;-B_{4},\;0)$,
$\hat{Q}_{5}^c\sim(\bar{3},\;2,\;-1/6,\;-(1+B_{4}),\;0)$, $\hat{U}_{5}\sim(3,\;1,\;2/3,\;1+B_{4},\;0)$,
$\hat{D}_{5}\sim(3,\;1,\;-1/3,\;1+B_{4},\;0)$) are respectively introduced.
 The detection of the lightest CP even Higgs $h^0$ at LHC\cite{Higgs} makes people to be convinced of the
Higgs mechanism.
To break lepton number and baryon number spontaneously, the Higgs superfields $\hat{\Phi}_{L},\hat{\varphi}_{L}$  and
$\hat{\Phi}_{B},\hat{\varphi}_{B}$ are introduced respectively, and they acquire nonzero vacuum expectation values (VEVs).
 The exotic quarks are very heavy and unstable. So the
superfields $\hat{X}$,
$\hat{X}^\prime$ are also introduced in the BLMSSM and the lightest superfields X can be a candidate for dark matter.

 The superpotential of BLMSSM is\cite{weBLMSSM}
\begin{eqnarray}
&&{\cal W}_{{BLMSSM}}={\cal W}_{{MSSM}}+{\cal W}_{B}+{\cal W}_{L}+{\cal W}_{X}\;,
\label{superpotential1}
\nonumber\\&&{\cal W}_{B}=\lambda_{Q}\hat{Q}_{4}\hat{Q}_{5}^c\hat{\Phi}_{B}+\lambda_{U}\hat{U}_{4}^c\hat{U}_{5}
\hat{\varphi}_{B}+\lambda_{D}\hat{D}_{4}^c\hat{D}_{5}\hat{\varphi}_{B}+\mu_{B}\hat{\Phi}_{B}\hat{\varphi}_{B}
\nonumber\\
&&\hspace{1.2cm}
+Y_{{u_4}}\hat{Q}_{4}\hat{H}_{u}\hat{U}_{4}^c+Y_{{d_4}}\hat{Q}_{4}\hat{H}_{d}\hat{D}_{4}^c
+Y_{{u_5}}\hat{Q}_{5}^c\hat{H}_{d}\hat{U}_{5}+Y_{{d_5}}\hat{Q}_{5}^c\hat{H}_{u}\hat{D}_{5}\;,
\nonumber\\
&&{\cal W}_{L}=Y_{{e_4}}\hat{L}_{4}\hat{H}_{d}\hat{E}_{4}^c+Y_{{\nu_4}}\hat{L}_{4}\hat{H}_{u}\hat{N}_{4}^c
+Y_{{e_5}}\hat{L}_{5}^c\hat{H}_{u}\hat{E}_{5}+Y_{{\nu_5}}\hat{L}_{5}^c\hat{H}_{d}\hat{N}_{5}
\nonumber\\
&&\hspace{1.2cm}
+Y_{\nu}\hat{L}\hat{H}_{u}\hat{N}^c+\lambda_{{N^c}}\hat{N}^c\hat{N}^c\hat{\varphi}_{L}
+\mu_{L}\hat{\Phi}_{L}\hat{\varphi}_{L}\;,
\nonumber\\
&&{\cal W}_{X}=\lambda_1\hat{Q}\hat{Q}_{5}^c\hat{X}+\lambda_2\hat{U}^c\hat{U}_{5}\hat{X}^\prime
+\lambda_3\hat{D}^c\hat{D}_{5}\hat{X}^\prime+\mu_{X}\hat{X}\hat{X}^\prime\;.
\label{superpotential-BL}
\end{eqnarray}
where ${\cal W}_{{MSSM}}$ is the superpotential of the MSSM.
The soft breaking terms $\mathcal{L}_{{soft}}$ of the BLMSSM can be written in the following form\cite{weBLMSSM}.
\begin{eqnarray}
&&{\cal L}_{{soft}}={\cal L}_{{soft}}^{MSSM}-(m_{{\tilde{\nu}^c}}^2)_{{IJ}}\tilde{N}_I^{c*}\tilde{N}_J^c
-m_{{\tilde{Q}_4}}^2\tilde{Q}_{4}^\dagger\tilde{Q}_{4}-m_{{\tilde{U}_4}}^2\tilde{U}_{4}^{c*}\tilde{U}_{4}^c
-m_{{\tilde{D}_4}}^2\tilde{D}_{4}^{c*}\tilde{D}_{4}^c
\nonumber\\
&&\hspace{1.3cm}
-m_{{\tilde{Q}_5}}^2\tilde{Q}_{5}^{c\dagger}\tilde{Q}_{5}^c-m_{{\tilde{U}_5}}^2\tilde{U}_{5}^*\tilde{U}_{5}
-m_{{\tilde{D}_5}}^2\tilde{D}_{5}^*\tilde{D}_{5}-m_{{\tilde{L}_4}}^2\tilde{L}_{4}^\dagger\tilde{L}_{4}
-m_{{\tilde{\nu}_4}}^2\tilde{N}_{4}^{c*}\tilde{N}_{4}^c
\nonumber\\
&&\hspace{1.3cm}
-m_{{\tilde{e}_4}}^2\tilde{E}_{_4}^{c*}\tilde{E}_{4}^c-m_{{\tilde{L}_5}}^2\tilde{L}_{5}^{c\dagger}\tilde{L}_{5}^c
-m_{{\tilde{\nu}_5}}^2\tilde{N}_{5}^*\tilde{N}_{5}-m_{{\tilde{e}_5}}^2\tilde{E}_{5}^*\tilde{E}_{5}
-m_{{\Phi_{B}}}^2\Phi_{B}^*\Phi_{B}
\nonumber\\
&&\hspace{1.3cm}
-m_{{\varphi_{B}}}^2\varphi_{B}^*\varphi_{B}-m_{{\Phi_{L}}}^2\Phi_{L}^*\Phi_{L}
-m_{{\varphi_{L}}}^2\varphi_{L}^*\varphi_{L}-\Big(m_{B}\lambda_{B}\lambda_{B}
+m_{L}\lambda_{L}\lambda_{L}+h.c.\Big)
\nonumber\\
&&\hspace{1.3cm}
+\Big\{A_{{u_4}}Y_{{u_4}}\tilde{Q}_{4}H_{u}\tilde{U}_{4}^c+A_{{d_4}}Y_{{d_4}}\tilde{Q}_{4}H_{d}\tilde{D}_{4}^c
+A_{{u_5}}Y_{{u_5}}\tilde{Q}_{5}^cH_{d}\tilde{U}_{5}+A_{{d_5}}Y_{{d_5}}\tilde{Q}_{5}^cH_{u}\tilde{D}_{5}
\nonumber\\
&&\hspace{1.3cm}
+A_{{BQ}}\lambda_{Q}\tilde{Q}_{4}\tilde{Q}_{5}^c\Phi_{B}+A_{{BU}}\lambda_{U}\tilde{U}_{4}^c\tilde{U}_{5}\varphi_{B}
+A_{{BD}}\lambda_{D}\tilde{D}_{4}^c\tilde{D}_{5}\varphi_{B}+B_{B}\mu_{B}\Phi_{B}\varphi_{B}
+h.c.\Big\}
\nonumber\\
&&\hspace{1.3cm}
+\Big\{A_{{e_4}}Y_{{e_4}}\tilde{L}_{4}H_{d}\tilde{E}_{4}^c+A_{{\nu_4}}Y_{{\nu_4}}\tilde{L}_{4}H_{u}\tilde{N}_{4}^c
+A_{{e_5}}Y_{{e_5}}\tilde{L}_{5}^cH_{u}\tilde{E}_{5}+A_{{\nu_5}}Y_{{\nu_5}}\tilde{L}_{5}^cH_{d}\tilde{N}_{5}
\nonumber\\
&&\hspace{1.3cm}
+A_{N}Y_{\nu}\tilde{L}H_{u}\tilde{N}^c+A_{{N^c}}\lambda_{{N^c}}\tilde{N}^c\tilde{N}^c\varphi_{L}
+B_{L}\mu_{L}\Phi_{L}\varphi_{L}+h.c.\Big\}
\nonumber\\
&&\hspace{1.3cm}
+\Big\{A_1\lambda_1\tilde{Q}\tilde{Q}_{5}^cX+A_2\lambda_2\tilde{U}^c\tilde{U}_{5}X^\prime
+A_3\lambda_3\tilde{D}^c\tilde{D}_{5}X^\prime+B_{X}\mu_{X}XX^\prime+h.c.\Big\}\;,
\label{soft-breaking}
\end{eqnarray}

 The $SU(2)_L$ singlets $\Phi_{B},\;\varphi_{B},\;\Phi_{L},\;
\varphi_{L}$ and the $SU(2)_L$ doublets $H_{u},\;H_{d}$ should obtain nonzero VEVs
 $\upsilon_{{B}},\;\overline{\upsilon}_{{B}},\;\upsilon_{L},\;\overline{\upsilon}_{L}$
and $\upsilon_{u},\;\upsilon_{d}$ respectively. Therefore, the local gauge symmetry $SU(2)_{L}\otimes U(1)_{Y}\otimes U(1)_{B}\otimes U(1)_{L}$
breaks down to the electromagnetic symmetry $U(1)_{e}$.
\begin{eqnarray}
&&H_{u}=\left(\begin{array}{c}H_{u}^+\\{1\over\sqrt{2}}\Big(\upsilon_{u}+H_{u}^0+iP_{u}^0\Big)\end{array}\right)\;,~~~~
H_{d}=\left(\begin{array}{c}{1\over\sqrt{2}}\Big(\upsilon_{d}+H_{d}^0+iP_{d}^0\Big)\\H_{d}^-\end{array}\right)\;,
\nonumber\\
&&\Phi_{B}={1\over\sqrt{2}}\Big(\upsilon_{B}+\Phi_{B}^0+iP_{B}^0\Big)\;,~~~~~~~~~
\varphi_{B}={1\over\sqrt{2}}\Big(\overline{\upsilon}_{B}+\varphi_{B}^0+i\overline{P}_{B}^0\Big)\;,
\nonumber\\
&&\Phi_{L}={1\over\sqrt{2}}\Big(\upsilon_{L}+\Phi_{L}^0+iP_{L}^0\Big)\;,~~~~~~~~~~
\varphi_{L}={1\over\sqrt{2}}\Big(\overline{\upsilon}_{L}+\varphi_{L}^0+i\overline{P}_{L}^0\Big)\;,
\label{VEVs}
\end{eqnarray}
In Ref.\cite{weBLMSSM}, the mass matrixes of Higgs, exotic quarks and exotic scalar quarks are obtained.
Some mass matrixes of exotic scalar leptons are discussed by the authors \cite{J.M. Arnold}.
Here, we show the mass matrixes of exotic scalar leptons in our notation. Because the super fields $\hat{N}^c$
are introduced in BLMSSM, the neutrinos can have tiny masses, and the scalar neutrinos are double as those in MSSM.
 \subsection{The mass matrix}
 After symmetry breaking the mass
matrix for neutrinos in the left-handed basis $(\nu,N^c)$
is given by the following matrix.
    \begin{eqnarray}
-\mathcal{L}_{mass}^{\nu}=(\bar{\nu}^I_R,\bar{N}^{cI}_R )\left(\begin{array}{cc}
  0&\frac{v_u}{\sqrt{2}}(Y_{\nu})^{IJ} \\
   \frac{v_u}{\sqrt{2}}(Y^{T}_{\nu})^{IJ}  & \frac{\bar{v}_L}{\sqrt{2}}(\lambda_{N^c})^{IJ}
    \end{array}\right) \left(\begin{array}{c}
   \nu^J_L\\   N^{cJ}_L
    \end{array}\right)+h.c.
      \end{eqnarray}
      Using the unitary transformations
      \begin{eqnarray}
&&\left(\begin{array}{l}\nu_{1L}^I\\\nu_{2L}^I\end{array}\right)
=U_{\nu^{IJ}}^\dagger\left(\begin{array}{c}
   \nu^J_L\\   N^{cJ}_L
    \end{array}\right)\;,\;\;
\left(\begin{array}{l}\nu_{1R}^I\\\nu_{2R}^I\end{array}\right)
=W_{\nu^{IJ}}^\dagger\left(\begin{array}{c}
   \nu^J_R\\ N^{cJ}_R
    \end{array}\right),
\end{eqnarray}
we diagonalize the mass matrix for neutrinos:
      \begin{eqnarray}
W_{\nu^{IJ}}^{\dag}\left(\begin{array}{cc}
  0&\frac{v_u}{\sqrt{2}}(Y_{\nu})^{IJ} \\
   \frac{v_u}{\sqrt{2}}(Y^{T}_{\nu})^{IJ}  & \frac{\bar{v}_L}{\sqrt{2}}(\lambda_{N^c})^{IJ}
    \end{array}\right) U_{\nu^{IJ}}=diag(m_{\nu_1^I},m_{\nu_2^I}).
      \end{eqnarray}
In a similar way, we obtain the exotic neutrinos mass matrix.
\begin{eqnarray}
-\mathcal{L}_{mass}^{\nu_{4,5}}=(\bar{N}_{4R},\bar{N}_{5R})\left(\begin{array}{cc}
  0&-\frac{v_d}{\sqrt{2}}Y_{\nu_5} \\
  \frac{v_u}{\sqrt{2}}Y_{\nu_4}  & 0
    \end{array}\right)\left(\begin{array}{c}
  N_{4L}\\   N_{5L}
    \end{array}\right)+h.c.
      \end{eqnarray}
Adopting the unitary transformations
      \begin{eqnarray}
&&\left(\begin{array}{l}N_{{4L}}^\prime\\N_{{5L}}^\prime\end{array}\right)
=U_{{N}}^\dagger\left(\begin{array}{l}N_{{4L}}\\N_{{5L}}\end{array}\right)\;,\;\;
\left(\begin{array}{l}N_{{4R}}^\prime\\N_{{5R}}^\prime\end{array}\right)
=W_{{N}}^\dagger\left(\begin{array}{l}N_{{4R}}\\N_{{5R}}\end{array}\right)\;,
\label{Qmixing-2/3-a}
\end{eqnarray}
the mass matrix of exotic neutrinos are diagonalized as
      \begin{eqnarray}
W_N^{\dag}\left(\begin{array}{cc}
  0&-\frac{v_d}{\sqrt{2}}Y_{\nu_5} \\
  \frac{v_u}{\sqrt{2}}Y_{\nu_4}  & 0
    \end{array}\right)  U_N=diag(m_{\nu_4},m_{\nu_5}).
      \end{eqnarray}
The mass matrix of exotic charged lepton are shown here
    \begin{eqnarray}
-\mathcal{L}_{mass}^{L_{4,5}}=(\bar{L}_{4R},\bar{L}_{5R})\left(\begin{array}{cc}
  0&\frac{v_u}{\sqrt{2}}Y_{e_5} \\
  -\frac{v_d}{\sqrt{2}}Y_{e_4}  & 0
    \end{array}\right)\left(\begin{array}{c}
  L_{4L}\\   L_{5L}
    \end{array}\right)+h.c.
      \end{eqnarray}
With the unitary transformations
\begin{eqnarray}
&&\left(\begin{array}{l}L_{{4L}}^\prime\\L_{{5L}}^\prime\end{array}\right)
=U_{{L}}^\dagger\cdot\left(\begin{array}{l}L_{{4L}}\\L_{{5L}}\end{array}\right)\;,\;\;
\left(\begin{array}{l}L_{{4R}}^\prime\\L_{{5R}}^\prime\end{array}\right)
=W_{{L}}^\dagger\cdot\left(\begin{array}{l}L_{{4R}}\\L_{{5R}}\end{array}\right)\;,
\label{Qmixing-2/3-a}
\end{eqnarray}
one can diagonalize the mass matrix of exotic charged lepton as
\begin{eqnarray}
W_L^{\dag}\left(\begin{array}{cc}
  0&\frac{v_u}{\sqrt{2}}Y_{e_5} \\
  -\frac{v_d}{\sqrt{2}}Y_{e_4}  & 0
    \end{array}\right)  U_L=diag(m_{l_4},m_{l_5})
      \end{eqnarray}
 From the superpotential and the soft breaking terms in BLMSSM Eq.(\ref{superpotential-BL}), the mass squared matrices of the scalar neutrinos
 and scalar exotic charged leptons are obtained.
 \begin{eqnarray}
&&-\mathcal{L} _{S}^{mass}=\tilde{n}^{\dagger}\cdot
{\cal M}_{\tilde{n}}^2\cdot\tilde{n}
+\tilde{n}^{\dagger}_4\cdot
{\cal M}_{\tilde{n}_4}^2\cdot\tilde{n}_4+\tilde{n}^{\dagger}_5\cdot
{\cal M}_{\tilde{n}_5}^2\cdot\tilde{n}_5\nonumber\\&&\hspace{2.0cm}+\tilde{e}^{\dagger}_4\cdot
{\cal M}_{\tilde{e}_4}^2\cdot\tilde{e}_4+\tilde{e}^{\dagger}_5\cdot
{\cal M}_{\tilde{e}_5}^2\cdot\tilde{e}_5
\label{SQmass-2/3}
\end{eqnarray}
 with $\tilde{n}^{T}=(\tilde{\nu}^{I},\tilde{N}^{cI*})$, $\tilde{n}_4^T=(\tilde{N}_{4},\tilde{N}_{4}^{c*})$
  , $\tilde{e}_4^T=(\tilde{E}_{4},\tilde{E}_{4}^{c*})$, $\tilde{n}_5^T=(\tilde{N}_{5},\tilde{N}_{5}^{c*})$
  and $\tilde{e}_5^T=(\tilde{E}_{5},\tilde{E}_{5}^{c*})$.
  The concrete forms for the mass squared matrices ${\cal M}_{\tilde{n}},{\cal M}_{\tilde{n}_4},{\cal M}_{\tilde{e}_4}
  ,{\cal M}_{\tilde{n}_5}$ and ${\cal M}_{\tilde{e}_5}$ are collected here.

The scalar neutrinos are enlarged by the superfields $\tilde{N}^c$ and the mass squared matrix reads as
    \begin{eqnarray}
  && {\cal M}^2_{\tilde{n}}(\tilde{\nu}_{I}^*\tilde{\nu}_{J})=\frac{g_1^2+g_2^2}{8}(v_d^2-v_u^2)\delta_{IJ}+g_L^2(\overline{v}^2_L-v^2_L)\delta_{IJ}
   +\frac{v_u^2}{2}(Y^\dag_{\nu}Y_\nu)_{IJ}+(M^2_{\tilde{L}})_{IJ},\nonumber\\&&
   {\cal M}^2_{\tilde{n}}(\tilde{N}_I^{c*}\tilde{N}_J^c)=-g_L^2(\overline{v}^2_L-v^2_L)\delta_{IJ}
   +\frac{v_u^2}{2}(Y^\dag_{\nu}Y_\nu)_{IJ}+2\overline{v}^2_L(\lambda_{N_c}^\dag\lambda_{N_c})_{IJ}\nonumber\\&&
   \hspace{1.8cm}+(M^2_{\tilde{\nu}})_{IJ}+\mu_L\frac{v_L}{\sqrt{2}}(\lambda_{N_c})_{IJ}
   -\frac{\overline{v}_L}{\sqrt{2}}(A_{N_c})_{IJ},\nonumber\\&&
   {\cal M}^2_{\tilde{n}}(\tilde{\nu}_I\tilde{N}_J^c)=\mu^*\frac{v_d}{\sqrt{2}}(Y_{\nu})_{IJ}-v_u\overline{v}_L(Y_{\nu}^\dag\lambda_{N_c})_{IJ}
   +\frac{v_u}{\sqrt{2}}(A_{N})_{IJ}.
   \end{eqnarray}
   The mass squared matrix of the 4th generation scalar neutrinos is
   \begin{eqnarray}
  &&  {\cal M}^2_{\tilde{n}_4}(\tilde{N}_4^*\tilde{N}_4)=\frac{g_1^2+g_2^2}{8}(v_d^2-v_u^2)+g_L^2L_4(\overline{v}^2_L-v^2_L)
   +\frac{v_u^2}{2}|Y_{\nu_4}|^2+M^2_{\tilde{L}_4},\nonumber\\&&
    {\cal M}^2_{\tilde{n}_4}(\tilde{N}_4^{c*}\tilde{N}_4^c)=-g_L^2L_4(\overline{v}^2_L-v^2_L)
   +\frac{v_u^2}{2}|Y_{\nu_4}|^2+M^2_{\tilde{\nu}_4},\nonumber\\&&
    {\cal M}^2_{\tilde{n}_4}(\tilde{N}_4\tilde{N}_4^c)=\mu^*\frac{v_d}{\sqrt{2}}Y_{\nu_4}+A_{\nu_4}\frac{v_u}{\sqrt{2}}.
   \end{eqnarray}
  The mass squared matrix of the 4th generation scalar charged leptons is
    \begin{eqnarray}
  &&  {\cal M}^2_{\tilde{e}_4}(\tilde{E}_{4}^*\tilde{E}_{4})=\frac{g_1^2-g_2^2}{8}(v_d^2-v_u^2)+g_L^2L_4(\overline{v}^2_L-v^2_L)
   +\frac{v_d^2}{2}|Y_{e_4}|^2+M^2_{\tilde{L}_4},\nonumber\\&&
   {\cal M}^2_{\tilde{e}_4}(\tilde{E}_4^{c*}\tilde{E}_4^c)=-\frac{g_1^2}{4}(v_d^2-v_u^2)-g_L^2L_4(\overline{v}^2_L-v^2_L)
   +\frac{v_d^2}{2}|Y_{e_4}|^2+M^2_{\tilde{e}_4}\nonumber\\&&
   {\cal M}^2_{\tilde{e}_4}(\tilde{E}_4\tilde{E}_4^c)=\mu^*\frac{v_u}{\sqrt{2}}Y_{e_4}+A_{e_4}\frac{v_d}{\sqrt{2}}.
   \end{eqnarray}
   The mass squared matrix of the 5th generation scalar neutrinos is
     \begin{eqnarray}
  &&{\cal M}^2_{\tilde{n}_5}(\tilde{N}_5^{c*}\tilde{N}_5^c)=-\frac{g_1^2+g_2^2}{8}(v_d^2-v_u^2)-g_L^2(3+L_4)(\overline{v}^2_L-v^2_L)
   +\frac{v_d^2}{2}|Y_{\nu_5}|^2+M^2_{\tilde{L}_5},\nonumber\\&&
   {\cal M}^2_{\tilde{n}_5}(\tilde{N}_5^{*}\tilde{N}_5)=g_L^2(3+L_4)(\overline{v}^2_L-v^2_L)
   +\frac{v_d^2}{2}|Y_{\nu_5}|^2+M^2_{\tilde{\nu}_5}\nonumber\\&&
   {\cal M}^2_{\tilde{n}_5}(\tilde{N}_5\tilde{N}_5^c)=\mu^*\frac{v_u}{\sqrt{2}}Y_{\nu_5}+A_{\nu_5}\frac{v_d}{\sqrt{2}}.
   \end{eqnarray}
  The mass squared matrix of the 5th generation scalar charged leptons is
    \begin{eqnarray}
  && {\cal M}^2_{\tilde{e}_5}(\tilde{E}_{5}^{c*}\tilde{E}_{5}^c)=-\frac{g_1^2-g_2^2}{8}(v_d^2-v_u^2)-g_L^2(3+L_4)(\overline{v}^2_L-v^2_L)
   +\frac{v_u^2}{2}|Y_{e_5}|^2+M^2_{\tilde{L}_5},\nonumber\\&&
   {\cal M}^2_{\tilde{e}_5}(\tilde{E}_5^{*}\tilde{E}_5)=\frac{g_1^2}{4}(v_d^2-v_u^2)+g_L^2(3+L_4)(\overline{v}^2_L-v^2_L)
   +\frac{v_u^2}{2}|Y_{e_5}|^2+M^2_{\tilde{e}_5}\nonumber\\&&
   {\cal M}^2_{\tilde{e}_5}(\tilde{E}_5\tilde{E}_5^c)=\mu^*\frac{v_d}{\sqrt{2}}Y_{e_5}+A_{e_5}\frac{v_u}{\sqrt{2}}.
   \end{eqnarray}

\subsection{The needed couplings}
We deduce the couplings between the charged Higgs and the exotic leptons(4,5) from the super potential in Eq.(\ref{superpotential-BL}).
  \begin{eqnarray}
  &&\mathcal{L}_{H^{\pm}N^\prime L^\prime}=\sum_{i,j=1}^2\overline{N}^\prime_{i+3}\Big((Y_{e_4}^*(U_{N}^\dag)^{i1}W_L^{2j}+Y_{\nu_5}^*(U_N^\dag)^{i2}W_{L}^{1j})\cos\beta\omega_+\nonumber\\&&
-(Y_{\nu_4}(W_{N}^\dag)^{i2}U_L^{1j}+Y_{e_5}(W_N^\dag)^{i1}U_{L}^{2j})\sin\beta\omega_-\Big)L^\prime_{j+3}G^{+}\nonumber\\&&+
\sum_{i,j=1}^2\overline{N}^\prime_{i+3}\Big(-(Y_{e_4}^*(U_{N}^\dag)^{i1}W_L^{2j}+Y_{\nu_5}^*(U_N^\dag)^{i2}W_{L}^{1j})\sin\beta\omega_+\nonumber\\&&
-(Y_{\nu_4}(W_{N}^\dag)^{i2}U_L^{1j}+Y_{e_5}(W_N^\dag)^{i1}U_{L}^{2j})\cos\beta\omega_-\Big)L^\prime_{j+3}H^{+}+h.c. \label{HpmLL}
  \end{eqnarray}
The couplings between neutral CP-even Higgs and the exotic leptons(4,5) are shown here.
  \begin{eqnarray}
  &&\mathcal{L}_{H^{0}L^\prime L^\prime}=\sum_{i,j=1}^2(-\frac{Y_{e_4}}{\sqrt{2}}(W_{L}^\dag)^{i2}U_L^{1j}\cos\alpha+\frac{Y_{e_5}}
  {\sqrt{2}}(W_{L}^\dag)^{i1}U_L^{2j}\sin\alpha)\overline{L}^\prime_{i+3}\omega_-L_{j+3}^\prime H^0
  \nonumber\\&&\hspace{1.6cm}+\sum_{i,j=1}^2(-\frac{Y_{e_4}^*}{\sqrt{2}}W_{L}^{2j}(U_L^\dag)^{i1}\cos\alpha+\frac{Y_{e_5}^*}{\sqrt{2}}W_{L}^{1j}(U_L^\dag)^{i2}\sin\alpha)
  \overline{L}_{i+3}^\prime\omega_+L_{j+3}^\prime H^0
  \nonumber\\&&\hspace{1.6cm}+\sum_{i,j=1}^2
  (\frac{Y_{e_4}}{\sqrt{2}}(W_{L}^\dag)^{i2}U_L^{1j}\sin\alpha+\frac{Y_{e_5}}{\sqrt{2}}(W_{L}^\dag)^{i1}U_L^{2j}\cos\alpha)\overline{L}^{\prime}_{i+3}\omega_-L^\prime_{j+3}h^0
  \nonumber\\&&\hspace{1.6cm}+\sum_{i,j=1}^2(\frac{Y_{e_4}^*}{\sqrt{2}}W_{L}^{2j}(U_L^\dag)^{i1}\sin\alpha+\frac{Y_{e_5}^*}{\sqrt{2}}W_{L}^{1j}(U_L^\dag)^{i2}\cos\alpha)
  \overline{L}^\prime_{i+3}\omega_+L^\prime_{j+3}h^0.\label{HLL}\\&&
 \mathcal{L}_{H^{0}N^\prime N^\prime}=\sum_{i,j=1}^2
  (-\frac{Y_{\nu_5}}{\sqrt{2}}(W_{N}^\dag)^{i1}U_N^{2j}\cos\alpha+\frac{Y_{\nu_4}}{\sqrt{2}}(W_{N}^\dag)^{i2}U_N^{1j}\sin\alpha)\overline{N}^\prime_{i+3}\omega_-N^\prime_{j+3}H^0
  \nonumber\\&&\hspace{1.6cm}+
  \sum_{i,j=1}^2(-\frac{Y_{\nu_5}^*}{\sqrt{2}}W_{N}^{1j}(U_N^\dag)^{i2}\cos\alpha+\frac{Y_{\nu_4}^*}{\sqrt{2}}W_{N}^{2j}(U_N^\dag)^{i1}\sin\alpha)
  \overline{N}^\prime_{i+3}\omega_+N^\prime_{j+3}H^0
  \nonumber\\&&\hspace{1.6cm}+\sum_{i,j=1}^2
  (\frac{Y_{\nu_5}}{\sqrt{2}}(W_{N}^\dag)^{i1}U_N^{2j}\sin\alpha+\frac{Y_{\nu_4}}{\sqrt{2}}(W_{N}^\dag)^{i2}U_N^{1j}\cos\alpha)\overline{N}^\prime_{i+3}
  \omega_-N^\prime_{j+3}h^0\nonumber\\&&\hspace{1.6cm}+
  \sum_{i,j=1}^2(\frac{Y_{\nu_5}^*}{\sqrt{2}}W_{N}^{1j}(U_N^\dag)^{i2}\sin\alpha+\frac{Y_{\nu_4}^*}{\sqrt{2}}W_{N}^{2j}(U_N^\dag)^{i1}\cos\alpha)
  \overline{N}^\prime_{i+3}\omega_+N^\prime_{j+3}h^0\label{ALL}.
  \end{eqnarray}

 Using the same method, we also get the couplings between neutral CP-odd Higgs and the exotic leptons(4,5).
  \begin{eqnarray}
  &&\mathcal{L}_{A^{0}L^\prime L^\prime}=\sum_{i,j=1}^2i(-\frac{Y_{e_4}}{\sqrt{2}}(W_{L}^\dag)^{i2}U_L^{1j}\cos\beta+\frac{Y_{e_5}}{\sqrt{2}}(W_{L}^\dag)^{i1}
  U_L^{2j}\sin\beta)\overline{L}^\prime_{i+3}\omega_-L^\prime_{j+3}G^0
  \nonumber\\&&\hspace{1.6cm}+\sum_{i,j=1}^2i(-\frac{Y_{e_4}^*}{\sqrt{2}}W_{L}^{2j}(U_L^\dag)^{i1}\cos\beta+\frac{Y_{e_5}^*}{\sqrt{2}}W_{L}^{1j}(U_L^\dag)^{i2}\sin\beta)
  \overline{L}^\prime_{i+3}\omega_+L^\prime_{j+3}G^0
  \nonumber\\&&\hspace{1.6cm}+\sum_{i,j=1}^2
  i(\frac{Y_{e_4}}{\sqrt{2}}(W_{L}^\dag)^{i2}U_L^{1j}\sin\beta+\frac{Y_{e_5}}{\sqrt{2}}(W_{L}^\dag)^{i1}U_L^{2j}\cos\beta)\overline{L}^\prime_{i+3}\omega_-L^{\prime}_{j+3}A^0
  \nonumber\\&&\hspace{1.6cm}+\sum_{i,j=1}^2i(\frac{Y_{e_4}^*}{\sqrt{2}}W_{L}^{2j}(U_L^\dag)^{i1}\sin\beta+\frac{Y_{e_5}^*}{\sqrt{2}}W_{L}^{1j}(U_L^\dag)^{i2}\cos\beta)
  \overline{L}^\prime_{i+3}\omega_+L^\prime_{j+3}A^0.
  \nonumber\\&&\mathcal{L}_{A^{0}N^\prime N^\prime}=
  \sum_{i,j=1}^2i(-\frac{Y_{\nu_5}}{\sqrt{2}}(W_{N}^\dag)^{i1}U_N^{2j}\cos\beta+\frac{Y_{\nu_4}}{\sqrt{2}}(W_{N}^\dag)^{i2}U_N^{1j}\sin\beta)\overline{N}^\prime_{i+3}
  \omega_-N^\prime_{j+3}G^0\nonumber\\&&\hspace{1.6cm}+\sum_{i,j=1}^2
  i(-\frac{Y_{\nu_5}^*}{\sqrt{2}}W_{N}^{1j}(U_N^\dag)^{i2}\cos\beta+\frac{Y_{\nu_4}^*}{\sqrt{2}}W_{N}^{2j}(U_N^\dag)^{i1}\sin\beta)
  \overline{N}^\prime_{i+3}\omega_+N^\prime_{j+3}G^0\nonumber\\&&\hspace{1.6cm}+\sum_{i,j=1}^2
  i(\frac{Y_{\nu_5}}{\sqrt{2}}(W_{N}^\dag)^{i1}U_N^{2j}\sin\beta+\frac{Y_{\nu_4}}{\sqrt{2}}(W_{N}^\dag)^{i2}U_N^{1j}\cos\beta)\overline{N}^\prime_{i+3}
  \omega_-N^\prime_{j+3}A^0\nonumber\\&&\hspace{1.6cm}+\sum_{i,j=1}^2
  i(\frac{Y_{\nu_5}^*}{\sqrt{2}}W_{N}^{1j}(U_N^\dag)^{i2}\sin\beta+\frac{Y_{\nu_4}^*}{\sqrt{2}}W_{N}^{2j}(U_N^\dag)^{i1}\cos\beta)
  \overline{N}^\prime_{i+3}\omega_+N^\prime_{j+3}A^0.
  \end{eqnarray}

 In the Barr-Zee type two-loop diagrams, the couplings between one vector boson and exotic leptons(4,5) are necessary.
  \begin{eqnarray}
  &&\mathcal{L}_{VL^\prime L^\prime}=\sum_{i,j=1}^2\Big[eF_{\mu}\overline{L}^\prime_{i+3}\gamma^{\mu}L^\prime_{i+3}
  -\frac{e}{2s_Wc_W}Z_{\mu}\overline{N}^\prime_{i+3}
  \Big( (U_{N}^\dag)^{i1}U_{N}^{1j}\gamma^{\mu}\omega_-+(W_{N}^\dag)^{i1}W_{N}^{1j}\gamma^{\mu}\omega_+\Big)N^\prime_{j+3}\nonumber\\&&
  +eZ_{\mu}\overline{L}_{i+3}
  \Big((-\frac{s_W}{c_W}\delta_{ij}
  +\frac{1}{2s_Wc_W}(U_{L}^\dag)^{i1}U_{L}^{1j})\gamma^{\mu}\omega_-+(-\frac{s_W}{c_W}\delta_{ij}
  +\frac{1}{2s_Wc_W}(W_{L}^\dag)^{i1}W_{L}^{1j})\gamma^{\mu}\omega_+\Big)L^\prime_{j+3}\nonumber\\&&
  \nonumber\\&&-\frac{e}{\sqrt{2}s_W}\overline{N}^\prime_{i+3}
  \Big( (U_{N}^\dag)^{i 1}U_{L}^{1j}\gamma^{\mu}\omega_--(W_{N}^\dag)^{i1}W_{L}^{1j}\gamma^{\mu}\omega_+\Big)L^\prime_{j+3}W_{\mu}^+\Big]+h.c.\label{VLL}
  \end{eqnarray}
  Here, we adopt the abbreviation notations $s_W=\sin\theta_W,~~c_W=\cos\theta_W$, where $\theta_W$ is the Weinberg angle.
 The exotic scalar leptons(4,5) have contributions to muon MDM at two-loop level. The couplings of one vector boson and exotic scalar leptons(4,5)
 are given out.
\begin{eqnarray}
&&\mathcal{L}_{V\tilde{L}^\prime \tilde{L}^\prime}=eF_{\mu}\sum_{i,j=1}^2\tilde{E}^{\prime i*}_4i\tilde{\partial}^{\mu}\tilde{E}^{\prime j}_4\delta^{ij}+
eZ_{\mu}\sum_{i,j=1}^2[-\frac{s_W}{c_W}\delta^{ij}+\frac{1}{2s_Wc_W}(Z_{\tilde{e}_4}^\dag)^{i1}Z_{\tilde{e}_4}^{1j}]\tilde{E}^{\prime i*}_4i\tilde{\partial}^{\mu}\tilde{E}^{\prime j}_4\nonumber\\&&
\hspace{1.4cm}+
eF_{\mu}\sum_{i,j=1}^2\tilde{E}^{\prime i*}_5i\tilde{\partial}^{\mu}\tilde{E}^{\prime j}_5\delta^{ij}+
eZ_{\mu}\sum_{i,j=1}^2[-\frac{s_W}{c_W}\delta^{ij}+\frac{1}{2s_Wc_W}(Z_{\tilde{e}_5}^\dag)^{i2}Z_{\tilde{e}_5}^{2j}]\tilde{E}^{\prime i*}_5i\tilde{\partial}^{\mu}\tilde{E}^{\prime j}_5\nonumber\\&&
\hspace{1.4cm}-\frac{e}{2s_Wc_W}Z_{\mu}\sum_{i,j=1}^2(Z_{\tilde{\nu}_4}^\dag)^{i1}Z_{\tilde{\nu}_4}^{1j}\tilde{N}^{\prime i*}_4i\tilde{\partial}^{\mu}\tilde{N}^{\prime j}_4
-\frac{e}{2s_Wc_W}Z_{\mu}\sum_{i,j=1}^2(Z_{\tilde{\nu}_5}^\dag)^{i2}Z_{\tilde{\nu}_5}^{2j}\tilde{N}^{\prime i*}_5i\tilde{\partial}^{\mu}\tilde{N}^{\prime j}_5
\nonumber\\&&
\hspace{1.4cm}-\frac{eW^+_{\mu}}{\sqrt{2}s_W}\sum_{i,j=1}^2(Z_{\tilde{\nu}_4}^\dag)^{i1}Z_{\tilde{e}_4}^{1j}\tilde{N}^{\prime i*}_4i\tilde{\partial}^{\mu}\tilde{E}^{\prime j}_4
+\frac{eW^+_{\mu}}{\sqrt{2}s_W}\sum_{i,j=1}^2(Z_{\tilde{\nu}_5}^\dag)^{i2}Z_{\tilde{e}_5}^{2j}\tilde{N}^{\prime i*}_5i\tilde{\partial}^{\mu}\tilde{E}^{\prime j}_5+h.c.\label{VSSLPLP}
\end{eqnarray}
 with $\tilde{\partial}^{\mu}=\overrightarrow{\partial}^{\mu}-\overleftarrow{\partial}^{\mu}$.
 $Z_{\tilde{\nu}_4},Z_{\tilde{e}_4},Z_{\tilde{\nu}_5},Z_{\tilde{e}_5}$ are the unitary
 matrices to diagonalize the mass squared matrices ${\cal M}^2_{\tilde{n}_4},{\cal M}^2_{\tilde{e}_4},{\cal M}^2_{\tilde{n}_5},{\cal M}^2_{\tilde{e}_5}$ respectively.
  \begin{eqnarray}
  &&Z_{\tilde{\nu}_4}^{\dag} {\cal M}^2_{\tilde{n}_4} Z_{\tilde{\nu}_4}=diag(m^2_{\tilde{\nu}^1_4},m^2_{\tilde{\nu}^2_4}),~~~
  Z_{\tilde{e}_4}^{\dag} {\cal M}^2_{\tilde{e}_4} Z_{\tilde{e}_4}=diag(m^2_{\tilde{e}^1_4},m^2_{\tilde{e}^2_4}),\nonumber\\&&
  Z_{\tilde{\nu}_5}^{\dag} {\cal M}^2_{\tilde{n}_5} Z_{\tilde{\nu}_5}=diag(m^2_{\tilde{\nu}^1_5},m^2_{\tilde{\nu}^2_5}),~~~
  Z_{\tilde{e}_5}^{\dag} {\cal M}^2_{\tilde{e}_5} Z_{\tilde{e}_5}=diag(m^2_{\tilde{e}^1_5},m^2_{\tilde{e}^2_5}).
  \end{eqnarray}
 For the couplings between vector Bosons and scalars, the VVSS type must be considered. Here, we just show the used
 coupling between $\gamma-V$ and two exotic scalar leptons(4,5).
\begin{eqnarray}
&&\mathcal{L}_{\gamma V\tilde{L}^\prime \tilde{L}^\prime}=\frac{e^2}{\sqrt{2}s_W}F^{\mu}W^+_{\mu}\sum_{i,j=1}^2\Big(-(Z_{\tilde{\nu}_4}^\dag)^{i1}
Z_{\tilde{e}_4}^{1j}\tilde{N}^{\prime i*}_4\tilde{E}^{\prime j}_4
+(Z_{\tilde{\nu}_5}^\dag)^{i2}Z_{\tilde{e}_5}^{2j}\tilde{N}^{\prime i*}_5\tilde{E}^{\prime j}_5\Big)\nonumber\\&&
+\frac{e^2}{s_Wc_W}F^{\mu}Z_{\mu}\sum_{i,j=1}^2\Big(((Z_{\tilde{e}_4}^\dag)^{i1}Z_{\tilde{e}_4}^{1j}-2s_W^2\delta_{ij})\tilde{E}^{\prime i*}_4\tilde{E}^{\prime j}_4
+((Z_{\tilde{e}_5}^\dag)^{i2}Z_{\tilde{e}_5}^{2j}-2s_W^2\delta_{ij})\tilde{E}^{\prime i*}_5\tilde{E}^{\prime j}_5\Big)\nonumber\\&&+
e^2F^{\mu}F_{\mu}\sum_{i,j=1}^2\delta_{ij}(\tilde{E}^{\prime i*}_4\tilde{E}^{\prime j}_4+\tilde{E}^{\prime i*}_5\tilde{E}^{\prime j}_5)+h.c.
\end{eqnarray}
The couplings between charged Higgs and exotic scalar leptons(4,5) are
 \begin{eqnarray}
  &&\mathcal{L}_{H^{\pm}\tilde{L}^\prime \tilde{L}^\prime}=\sum_{i,j=1}^2\tilde{N}^{\prime i*}_4\tilde{E}^{\prime j}_4G^+[(L^u_4)_{ij}\sin\beta+(L^d_4)_{ij}\cos\beta]\nonumber\\&&
  \hspace{1.6cm}+\sum_{i,j=1}^2\tilde{N}^{\prime i*}_4\tilde{E}^{\prime j}_4H^+[(L^u_4)_{ij}\cos\beta-(L^d_4)_{ij}\sin\beta]\nonumber\\ &&
  \hspace{1.6cm}+\sum_{i,j=1}^2\tilde{N}^{\prime i*}_5\tilde{E}^{\prime j}_5G^+[(L^u_5)_{ij}\sin\beta+(L^d_5)_{ij}\cos\beta]\nonumber\\ &&
  \hspace{1.6cm}+\sum_{i,j=1}^2\tilde{N}^{\prime i*}_5\tilde{E}^{\prime j}_5H^+[(L^u_5)_{ij}\cos\beta-(L^d_5)_{ij}\sin\beta]+h.c.\label{HPMSSLP}
\end{eqnarray}
with
\begin{eqnarray}
 &&(L^u_4)_{ij}=\frac{V_{EW}\sin\beta}{\sqrt{2}}(-\frac{e^2}{2s_W^2}+|Y_{\nu_4}|^2)(Z_{\tilde{\nu}_4}^\dag)^{i1}Z_{\tilde{e}_4}^{1j}
  -\mu Y_{e_4}^*(Z_{\tilde{\nu}_4}^\dag)^{i1}Z_{\tilde{e}_4}^{2j}\nonumber\\&&\hspace{1.6cm}-
  \frac{V_{EW}\cos\beta}{\sqrt{2}}Y_{e_4}^*Y_{\nu_4}(Z_{\tilde{\nu}_4}^\dag)^{i2}Z_{\tilde{e}_4}^{2j}
  +A_{N_4}(Z_{\tilde{\nu}_4}^\dag)^{i2}Z_{\tilde{e}_4}^{1j},\nonumber\\&&
  (L^d_4)_{ij}=\frac{V_{EW}\cos\beta}{\sqrt{2}}(-\frac{e^2}{2s_W^2}+|Y_{e_4}|^2)(Z_{\tilde{\nu}_4}^\dag)^{i1}Z_{\tilde{e}_4}^{1j}
  -\mu^* Y_{\nu_4}(Z_{\tilde{\nu}_4}^\dag)^{i2}Z_{\tilde{e}_4}^{1j}\nonumber\\&&\hspace{1.6cm}-
  \frac{V_{EW}\sin\beta}{\sqrt{2}}Y_{e_4}^*Y_{\nu_4}(Z_{\tilde{\nu}_4}^\dag)^{i2}Z_{\tilde{e}_4}^{2j}
  +A_{e_4}^*(Z_{\tilde{\nu}_4}^\dag)^{i1}Z_{\tilde{e}_4}^{2j},\nonumber\\&&
  (L^u_5)_{ij}=\frac{V_{EW}\sin\beta}{\sqrt{2}}(-\frac{e^2}{2s_W^2}+|Y_{e_5}|^2)(Z_{\tilde{\nu}_5}^\dag)^{i2}Z_{\tilde{e}_5}^{2j}
  -\mu Y_{\nu_5}^*(Z_{\tilde{\nu}_5}^\dag)^{i1}Z_{\tilde{e}_5}^{2j}\nonumber\\&&\hspace{1.6cm}-
  \frac{V_{EW}\cos\beta}{\sqrt{2}}Y_{e_5}Y_{\nu_5}^*(Z_{\tilde{\nu}_5}^\dag)
  ^{i1}Z_{\tilde{e}_5}^{1j}
  +A_{e_5}(Z_{\tilde{\nu}_5}^\dag)^{i2}Z_{\tilde{e}_5}^{1j},\nonumber\\&&
  (L^d_5)_{ij}=\frac{V_{EW}\cos\beta}{\sqrt{2}}(-\frac{e^2}{2s_W^2}+|Y_{e_5}|^2)(Z_{\tilde{\nu}_5}^\dag)^{i2}Z_{\tilde{e}_5}^{2j}
  -\mu^* Y_{e_5}(Z_{\tilde{\nu}_5}^\dag)^{i2}Z_{\tilde{e}_5}^{1j}\nonumber\\&&\hspace{1.6cm}
  -\frac{V_{EW}\sin\beta}{\sqrt{2}}Y_{e_5}Y_{\nu_5}^*(Z_{\tilde{\nu}_5}^\dag)^{i1}Z_{\tilde{e}_5}^{1j}
  +A_{N_5}^*(Z_{\tilde{\nu}_5}^\dag)^{i1}Z_{\tilde{e}_5}^{2j}.\label{HPMSSLP1}
  \end{eqnarray}

  The couplings between the neutral CP-even Higgs and the exotic scalar lepton(4,5) are also collected here.
 \begin{eqnarray}
  &&\mathcal{L}_{H^{0}\tilde{L}^\prime \tilde{L}^\prime}=\sum_{i,j=1}^2\tilde{N}^{\prime i*}_4\tilde{N}^{\prime j}_4
  \Big(H^0[(N^u_4)_{ij}\sin\alpha+(N^d_4)_{ij}\cos\alpha]
  +h^0[(N^u_4)_{ij}\cos\alpha-(N^d_4)_{ij}\sin\alpha]\Big)\nonumber\\&&\hspace{1.4cm}+\sum_{i,j=1}^2\tilde{N}^{\prime i*}_5\tilde{N}^{\prime j}_5
  \Big(H^0[(N^u_5)_{ij}\sin\alpha+(N^d_5)_{ij}\cos\alpha]
  +h^0[(N^u_5)_{ij}\cos\alpha-(N^d_5)_{ij}\sin\alpha]\Big)\nonumber\\&&\hspace{1.4cm}+\sum_{i,j=1}^2\tilde{E}^{\prime i*}_4\tilde{E}^{\prime j}_4
  \Big(H^0[(E^u_4)_{ij}\sin\alpha+(E^d_4)_{ij}\cos\alpha]
  +h^0[(E^u_4)_{ij}\cos\alpha-(E^d_4)_{ij}\sin\alpha]\Big)\nonumber\\&&\hspace{1.4cm}+\sum_{i,j=1}^2\tilde{E}^{\prime i*}_5\tilde{E}^{\prime j}_5
  \Big(H^0[(E^u_5)_{ij}\sin\alpha+(E^d_5)_{ij}\cos\alpha]
  +h^0[(E^u_5)_{ij}\cos\alpha-(E^d_5)_{ij}\sin\alpha]\Big),\label{LHLpLp}
  \end{eqnarray}
  where the concrete forms of the coupling constants $N^{u,d}_{4,5},E^{u,d}_{4,5}$ are
  \begin{eqnarray}
 &&(N^u_4)_{ij}=\frac{e^2}{4s_W^2c_W^2}V_{EW}\sin\beta (Z_{\tilde{\nu}_4}^\dag)^{i1}Z_{\tilde{\nu}_4}^{1j}-V_{EW}\sin\beta|Y_{\nu_4}|^2\delta_{ij}
  -\frac{A_{N_4}}{\sqrt{2}}(Z_{\tilde{\nu}_4}^\dag)^{i2}Z_{\tilde{\nu}_4}^{1j},\nonumber\\&&
  (N^d_4)_{ij}=-\frac{e^2}{4s_W^2c_W^2}V_{EW}\cos\beta (Z_{\tilde{\nu}_4}^\dag)^{i1}Z_{\tilde{\nu}_4}^{1j}-\frac{\mu^*}{\sqrt{2}}Y_{\nu_4}
  (Z_{\tilde{\nu}_4}^\dag)^{i2}Z_{\tilde{\nu}_4}^{1j},\nonumber\\&&(N^u_5)_{ij}=-\frac{e^2}{4s_W^2c_W^2}V_{EW}\sin\beta (Z_{\tilde{\nu}_5}^\dag)^{i2}Z_{\tilde{\nu}_5}^{2j}-\frac{\mu^*}{\sqrt{2}}Y_{\nu_5}
  (Z_{\tilde{\nu}_5}^\dag)^{i2}Z_{\tilde{\nu}_5}^{1j},\nonumber\\&&(N^d_5)_{ij}=\frac{e^2}{4s_W^2c_W^2}V_{EW}\cos\beta (Z_{\tilde{\nu}_5}^\dag)^{i2}Z_{\tilde{\nu}_5}^{2j}-V_{EW}\cos\beta|Y_{\nu_5}|^2\delta_{ij}
  -\frac{A_{N_5}}{\sqrt{2}}(Z_{\tilde{\nu}_5}^\dag)^{i2}Z_{\tilde{\nu}_5}^{1j},\label{LHLpLp1}
  \\&&
  (E^u_4)_{ij}=-e^2V_{EW}\sin\beta(\frac{1}{2c_W^2}\delta_{ij}+\frac{1-4s_W^2}{4s_W^2c_W^2}(Z_{\tilde{e}_4}^\dag)^{i1}Z_{\tilde{e}_4}^{1j})
  -\frac{\mu^*}{\sqrt{2}}Y_{e_4}(Z_{\tilde{e}_4}^\dag)^{i2}Z_{\tilde{e}_4}^{1j},\nonumber\\&&
  (E^d_4)_{ij}=e^2V_{EW}\cos\beta(\frac{1}{2c_W^2}\delta_{ij}+\frac{1-4s_W^2}{4s_W^2c_W^2}(Z_{\tilde{e}_4}^\dag)^{i1}Z_{\tilde{e}_4}^{1j})
  -V_{EW}\cos\beta|Y_{e_4}|^2\delta_{ij}-\frac{A_{E_4}}{\sqrt{2}}(Z_{\tilde{e}_4}^\dag)^{i2}Z_{\tilde{e}_4}^{1j},\nonumber\\&&
  (E^u_5)_{ij}=e^2V_{EW}\sin\beta(\frac{1}{2c_W^2}\delta_{ij}+\frac{1-4s_W^2}{4s_W^2c_W^2}(Z_{\tilde{e}_5}^\dag)^{i2}Z_{\tilde{e}_5}^{2j})
  -V_{EW}\sin\beta|Y_{e_5}|^2\delta_{ij}-\frac{A_{E_5}}{\sqrt{2}}(Z_{\tilde{e}_5}^\dag)^{i2}Z_{\tilde{e}_5}^{1j},\nonumber\\&&
  (E^d_5)_{ij}=-e^2V_{EW}\cos\beta(\frac{1}{2c_W^2}\delta_{ij}+\frac{1-4s_W^2}{4s_W^2c_W^2}(Z_{\tilde{e}_4}^\dag)^{i2}Z_{\tilde{e}_4}^{2j})
  -\frac{\mu^*}{\sqrt{2}}Y_{e_5}(Z_{\tilde{e}_5}^\dag)^{i2}Z_{\tilde{e}_5}^{1j}.\label{LHLpLp2}
  \end{eqnarray}
  Similarly, the couplings between the CP-odd Higgs and exotic scalar leptons(4,5) are obtained.
  {\small
  \begin{eqnarray}
  &&\mathcal{L}_{A^{0}\tilde{L}^\prime \tilde{L}^\prime}=-\frac{i}{\sqrt{2}}\sum_{i,j=1}^2
  \tilde{N}^{\prime i*}_4\tilde{N}^{\prime j}_4(\cos\beta A^0+\sin\beta G^0)A_{N_4}(Z_{\tilde{\nu}_4}^\dag)^{i2}Z_{\tilde{\nu}_4}^{1j}
  \nonumber\\&&\hspace{1.4cm}-\frac{i}{\sqrt{2}}\sum_{i,j=1}^2\tilde{N}^{\prime i*}_5\tilde{N}^{\prime j}_5(-\sin\beta A^0+\cos\beta G^0)A_{N_5}(Z_{\tilde{\nu}_5}^\dag)^{i2}Z_{\tilde{\nu}_5}^{1j}\nonumber\\&&\hspace{1.4cm}-\frac{i}{\sqrt{2}}\sum_{i,j=1}^2\tilde{E}^{\prime i*}_4\tilde{E}^{\prime j}_4(-\sin\beta A^0+\cos\beta G^0)A_{E_4}(Z_{\tilde{e}_4}^\dag)^{i2}Z_{\tilde{e}_4}^{1j}\nonumber\\&&\hspace{1.4cm}-\frac{i}{\sqrt{2}}\sum_{i,j=1}^2\tilde{E}^{\prime i*}_5\tilde{E}^{\prime j}_5  (\cos\beta A^0+\sin\beta G^0)A_{E_5}(Z_{\tilde{e}_5}^\dag)^{i2}Z_{\tilde{e}_5}^{1j}.\label{LALpLp}
  \end{eqnarray}}
The couplings between neutral Higgs and exotic quarks (scalar quarks) can be found in our previous work\cite{weBLMSSM}.
In Ref.\cite{smneutron}, the couplings between charged Higgs and exotic quarks are also given out. To complete the couplings, we deduce
the changed Higgs-exotic scalar quarks couplings.
  \begin{eqnarray}
  &&\mathcal{L}_{H^{\pm}\tilde{\mathcal{Q}}\tilde{\mathcal{Q}}}=\sum_{j,k=1}^4\tilde{\mathcal{U}}_{j}^*\tilde{\mathcal{D}}_k
  G^+[(R^u)_{jk}\sin\beta+(R^d)_{jk}\cos\beta]\nonumber\\&&
  \hspace{1.4cm}+\sum_{j,k=1}^4\tilde{\mathcal{U}}_{j}^*\tilde{\mathcal{D}}_kH^+[(R^u)_{jk}\cos\beta-(R^d)_{jk}\sin\beta]+h.c.\label{HPMSSQP}
  \end{eqnarray}
The concrete forms of the coupling constants $(R^u)_{jk},(R^d)_{jk}$ read as
  \begin{eqnarray}
  &&(R^u)_{jk}=-\frac{\sqrt{2}e^2}{4s_{_W}^2}v_u(U^{\dag}_{j1}D_{1k}+U^{\dag}_{j3}D_{3k})-\mu Y_{d_4}^*U^{\dag}_{j1}D_{2k}
  -\mu Y_{u_5}^*U^{\dag}_{j4}D_{3k}\nonumber\\&&\hspace{1.4cm}+\frac{v_B}{\sqrt{2}} \lambda_{Q}^*Y_{u_4}U^{\dag}_{j2}D_{3k}
  +\frac{v_d}{\sqrt{2}} Y_{u_4}Y_{d_4}^*U^{\dag}_{j2}D_{2k}+\frac{v_d}{\sqrt{2}} Y_{d_5}Y_{u_5}^*U^{\dag}_{j4}D_{4k}\nonumber\\&&
  \hspace{1.4cm}-\frac{v_B}{\sqrt{2}} Y_{d_5}\lambda_{Q}^*U^{\dag}_{j1}D_{4k}-\frac{\bar{v}_B}{\sqrt{2}} Y_{u_4}\lambda_{U}^*U^{\dag}_{j4}D_{1k}-\frac{\bar{v}_B}{\sqrt{2}} Y_{d_5}\lambda_{d}^*U^{\dag}_{j2}D_{2k}\nonumber\\&&\hspace{1.4cm}
  +A_{u_4}U^{\dag}_{j2}D_{1k}+A_{d_5}U^{\dag}_{j3}D_{4k},\nonumber\\&&
  (R^d)_{jk}=-\frac{\sqrt{2}e^2}{4s_{_W}^2}v_d(U^{\dag}_{j1}D_{1k}+U^{\dag}_{j3}D_{3k})-\mu^* Y_{u_4}U^{\dag}_{j2}D_{1k}
  -\mu^* Y_{d_5}U^{\dag}_{j3}D_{4k}\nonumber\\&&\hspace{1.4cm}-\frac{v_B}{\sqrt{2}} \lambda_{Q}Y_{d_4}^*U^{\dag}_{j3}D_{2k}
  +\frac{v_u}{\sqrt{2}} Y_{u_4}Y_{d_4}^*U^{\dag}_{j2}D_{2k}
  +\frac{v_B}{\sqrt{2}} \lambda_QY_{u_5}^*U^{\dag}_{j4}D_{1k}\nonumber\\&&\hspace{1.4cm}
  +\frac{v_u}{\sqrt{2}} Y_{d_5}Y_{u_5}^*U^{\dag}_{j4}D_{4k}
  -\frac{\bar{v}_B}{\sqrt{2}} Y_{d_4}^*\lambda_{D}U^{\dag}_{j1}D_{4k}-\frac{\bar{v}_B}{\sqrt{2}} Y_{u_5}^*\lambda_{U}U^{\dag}_{j2}D_{3k}\nonumber\\&&\hspace{1.4cm}
  +A_{d_4}^*U^{\dag}_{j1}D_{2k}+A_{u_5}^*U^{\dag}_{j4}D_{3k}.\label{HPMSSQP1}
  \end{eqnarray}
  One vector Boson can couple with the exotic scalar quarks
\begin{eqnarray}
&&\mathcal{L}_{V\tilde{\mathcal{Q}}\tilde{\mathcal{Q}}}=\frac{-e}{\sqrt{2}s_W}\sum_{j,\beta=1}^4
\Big(U^{\dag}_{j1}D_{1\beta}-U^{\dag}_{j3}D_{3\beta}\Big)W_{\mu}^+\tilde{\mathcal{U}}_{j}^*i\tilde{\partial}^{\mu}
\tilde{\mathcal{D}}_{\beta}-\frac{2}{3}e\sum_{j,\beta=1}^4\delta_{j\beta}F_{\mu}
\tilde{\mathcal{U}}_{j}^*i\tilde{\partial}^{\mu}
\tilde{\mathcal{U}}_{\beta}\nonumber\\&&+\frac{e}{3}\sum_{j,\beta=1}^4\delta_{j\beta}F_{\mu}
\tilde{\mathcal{D}}_{j}^*i\tilde{\partial}^{\mu}
\tilde{\mathcal{D}}_{\beta}+\frac{e}{6s_Wc_W}\sum_{j,\beta=1}^4
\Big(4s^2_W\delta_{j\beta}-3(U^{\dag}_{j1}U_{1\beta}+U^{\dag}_{j3}U_{3\beta})\Big)Z_{\mu}
\tilde{\mathcal{U}}_{j}^*i\tilde{\partial}^{\mu}
\tilde{\mathcal{U}}_{\beta}\nonumber\\&&+\frac{e}{6s_Wc_W}\sum_{j,\beta=1}^4
\Big(-2s^2_W\delta_{j\beta}+3(D^{\dag}_{j1}D_{1\beta}+D^{\dag}_{j3}D_{3\beta})\Big)Z_{\mu}
\tilde{\mathcal{D}}_{j}^*i\tilde{\partial}^{\mu}
\tilde{\mathcal{D}}_{\beta}
+h.c.\label{VSSQpQP}
\end{eqnarray}

The couplings between photon-vector boson-exotic scalar quarks must be taken into account.
\begin{eqnarray}
&&\mathcal{L}_{\gamma V\tilde{\mathcal{Q}}\tilde{\mathcal{Q}}}=\frac{e^2}{3\sqrt{2}s_W}\sum_{i,j}^4
\Big(U^{\dag}_{i1}D_{1j}-U^{\dag}_{i3}D_{3j}\Big)W_{\mu}^+F^{\mu}\tilde{\mathcal{U}}_{i}^*
\tilde{\mathcal{D}}_{j}+\frac{4e^2}{9}\sum_{i,j}^4
\delta_{ij}F_{\mu}F^{\mu}\tilde{\mathcal{U}}_{i}^*
\tilde{\mathcal{U}}_{j}\nonumber\\&&+\frac{e^2}{9}\sum_{i,j}^4
\delta_{ij}F_{\mu}F^{\mu}\tilde{\mathcal{D}}_{i}^*
\tilde{\mathcal{D}}_{j}+\frac{e^2}{9s_Wc_W}\sum_{i,j=1}^4
\Big(6(U^{\dag}_{i1}U_{1j}+U^{\dag}_{i3}U_{3j})-8s_W^2\delta_{ij}\Big)Z^{\mu}F_{\mu}
\tilde{\mathcal{U}}_{i}^*
\tilde{\mathcal{U}}_{j}\nonumber\\&&+\frac{e^2}{9s_Wc_W}\sum_{i,j=1}^4
\Big(3(D^{\dag}_{i1}D_{1j}+D^{\dag}_{i3}D_{3j})-2s_W^2\delta_{ij}\Big)Z^{\mu}F_{\mu}
\tilde{\mathcal{D}}_{i}^*
\tilde{\mathcal{D}}_{j}+h.c.
\end{eqnarray}
Because the exotic quark are very heavy, they can give considerable contribution to the muon MDM through the
coupling between Higgs and exotic quarks. We give out the coupling between vector Boson and exotic quarks.
\begin{eqnarray}
&&\mathcal{L}_{ V\mathcal{Q} \mathcal{Q}}=-\frac{2e}{3}F_{\mu}\sum_{i=1}^2
\bar{t}_{i+3}\gamma^{\mu}t_{i+3}+\frac{e}{3}F_{\mu}\sum_{i=1}^2\bar{b}_{i+3}\gamma^{\mu}b_{i+3}\nonumber
\\&&\hspace{1.4cm}+\frac{e}{6s_{W}c_{W}}Z_{\mu}\sum_{j,k=1}^2\bar{t}_{j+3}\Big[\Big((1-4c_{W}^2)\delta_{jk}
+3(U_t^{\dag})_{j2}(U_t)_{2k}\Big)\gamma^{\mu}\omega_-\nonumber\\&&\hspace{1.4cm}
+\Big((1-4c_{W}^2)\delta_{jk}
+3(W_t^{\dag})_{j2}(W_t)_{2k}\Big)\gamma^{\mu}\omega_+
\Big]t_{k+3}\nonumber\\&&\hspace{1.4cm}
+\frac{e}{6s_{W}c_{W}}Z_{\mu}\sum_{j,k=1}^2\bar{b}_{j+3}\Big[\Big((1+2c_{W}^2)\delta_{jk}
-3(U_b^{\dag})_{j2}(U_b)_{2k}\Big)\gamma^{\mu}\omega_-\nonumber\\&&\hspace{1.4cm}
+\Big((1+2c_{W}^2)\delta_{jk}
-3(W_b^{\dag})_{j2}(W_b)_{2k}\Big)\gamma^{\mu}\omega_+
\Big]b_{k+3}\nonumber\\&&\hspace{1.4cm}
+\frac{eW_{\mu}^+}{\sqrt{2}s_{W}}\sum_{j,k=1}^2\bar{t}_{j+3}\Big[
(W_t^{\dag})_{j1}(W_b)_{1k}\gamma^{\mu}\omega_+-(U_t^{\dag})_{j1}(U_b)_{1k}\gamma^{\mu}\omega_-
\Big]b_{k+3}+h.c.\label{VQQ}
\end{eqnarray}

\section{formulation}
 We use the effective Lagrangian method, the Feynman amplitude can be expressed by these dimension 6 operators.
\begin{eqnarray}
&&\mathcal{O}_1^{\mp}=\frac{1}{(4\pi)^2}\bar{l}(i\mathcal{D}\!\!\!\slash)^3\omega_{\mp}l,
\nonumber\\
&&\mathcal{O}_2^{\mp}=\frac{eQ_f}{(4\pi)^2}\overline{(i\mathcal{D}_{\mu}l)}\gamma^{\mu}
F\cdot\sigma\omega_{\mp}l,
\nonumber\\
&&\mathcal{O}_3^{\mp}=\frac{eQ_f}{(4\pi)^2}\bar{l}F\cdot\sigma\gamma^{\mu}
\omega_{\mp}(i\mathcal{D}_{\mu}l),\nonumber\\
&&\mathcal{O}_4^{\mp}=\frac{eQ_f}{(4\pi)^2}\bar{l}(\partial^{\mu}F_{\mu\nu})\gamma^{\nu}
\omega_{\mp}l,\nonumber\\&&
\mathcal{O}_5^{\mp}=\frac{m_l}{(4\pi)^2}\bar{l}(i\mathcal{D}\!\!\!\slash)^2\omega_{\mp}l,
\nonumber\\&&\mathcal{O}_6^{\mp}=\frac{eQ_fm_l}{(4\pi)^2}\bar{l}F\cdot\sigma
\omega_{\mp}l.
\end{eqnarray}

with $\mathcal{D}_{\mu}=\partial_{\mu}+ieA_{\mu}$ and $\omega_{\mp}=\frac{1\mp\gamma_5}{2}$. $F_{_{\mu\nu}}$ is the electromagnetic field strength, and
$m_{_l}$ is the lepton mass. Using the equations of motion to the incoming and out going leptons separately, only the
$\mathcal{O}_{2,3,6}^{\mp}$ contribute to lepton MDM and EDM. Therefore, the Wilson coefficients of the operators $\mathcal{O}_{2,3,6}^{\mp}$ in the effective Lagrangian are of interest and their dimensions are -2.
The lepton MDM is the combination of the Wilson coefficients $C^{\mp}_{2,3,6}$ and can be obtained from the following effective Lagrangian
\begin{eqnarray}
&&{\cal L}_{_{MDM}}={e\over4m_{_l}}\;a_{_l}\;\bar{l}\sigma^{\mu\nu}
l\;F_{_{\mu\nu}}\label{adm}.
\end{eqnarray}

\subsection{the one-loop corrections}
   In BLMSSM, the masses of the neutrinos, scalar neutrinos and scalar charged leptons are all
adopted comparing with those in MSSM. In BLMSSM, $\lambda_B$ (the superpartners of the new baryon boson) and $\psi_{\Phi_B},\psi_{\varphi_B}$
 (the superpartners of the $SU(2)_L$ singlets $\Phi_B,\varphi_B$) mix and generate three baryon neutralinos.
  Three lepton neutralinos are made up of $\lambda_L$ (the superpartners of the new lepton boson) and $\psi_{\Phi_L},\psi_{\varphi_L}$
 (the superpartners of the $SU(2)_L$ singlets $\Phi_L,\varphi_L$). There are also four MSSM neutralinos and they do not mix with
 baryon neutralinos and lepton neutralinos. That is to say the four MSSM neutralinos in BLMSSM are same as those in MSSM.
 Therefore in BLMSSM there are ten neutralinos, but three baryon neutralinos and three lepton neutralinos have none
 contribution to lepton MDM in our studied diagrams.

 The one loop new physics contributions to muon MDM, comes from the diagrams in Fig.1.
The one-loop triangle diagrams are obtained from the one-loop self-energy diagrams by attaching a photon on the internal
line in all possible ways. In BLMSSM, the one-loop corrections are similar to the MSSM results in
analytic form. The differences are: 1. The squared mass matrixes of scalar leptons because of
new parameters $g_L,\bar{v}_L,v_L$ and so on. 2. Right-handed neutrinos and scalar neutrinos
are introduced, which leads to the neutrinos and scalar neutrinos are doubled.
   The one-loop self-energy diagrams can be divided into four parts according to the virtual particles: 1. scalar neutrino-chargino; 2.
    neutral Higgs and lepton; 3. charged Higgs and neutrino; 4. scalar charged lepton and neutralino.
    \begin{figure}[h]
\setlength{\unitlength}{1mm}
\centering
\includegraphics[width=3.0in]{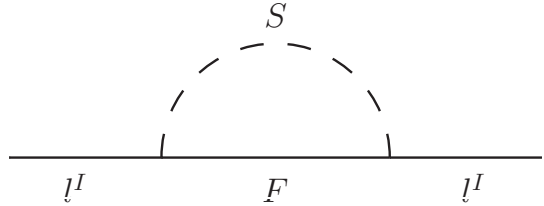}
\caption[]{The generic one loop self-energy diagram for lepton.}\label{oneloopfig}
\end{figure}
The lepton flavor mixing is also taken into account, whose contribution is considerable.
The corrections to muon MDM from neutralinos and scalar leptons  are expressed as
\begin{eqnarray}
&&a_{1}^{\tilde{L}\chi^{0}}=
-\frac{e^2}{2s_W^2}\sum_{i=1}^6\sum_{j=1}^4\Big[\texttt{Re}[(\mathcal{S}_1)^I_{ij}(\mathcal{S}_2)^{I*}_{ij}]
\sqrt{x_{\chi_j^{0}}x_{m_{l^I}}}x_{\tilde{L}_i}\frac{\partial^2 \mathcal{B}(x_{\chi_j^{0}},x_{\tilde{L}_i})}{\partial x_{\tilde{L}_i}^2}
\nonumber\\&&\hspace{1.4cm}+\frac{1}{3}(|(\mathcal{S}_1)^I_{ij}|^2+|(\mathcal{S}_2)^I_{ij}|^2)x_{\tilde{L}_i}x_{m_{l^I}}
\frac{\partial\mathcal{B}_1(x_{\chi_j^{0}},x_{\tilde{L}_i})}{\partial x_{\tilde{L}_i}}\Big],
\end{eqnarray}
where the couplings $(\mathcal{S}_1)^I_{ij},\;(\mathcal{S}_2)^{I}_{ij}$ are shown as
\begin{eqnarray}
&&(\mathcal{S}_1)^I_{ij}=\frac{1}{c_W}Z_{\tilde{L}}^{Ii*}(Z_N^{1j}s_W+Z_N^{2j}c_W)-\frac{m_{l^I}}{\cos\beta m_W}Z_{\tilde{L}}^{(I+3)i*}Z_N^{3j},\nonumber\\&&
(\mathcal{S}_2)^I_{ij}=-2\frac{s_W}{c_W}Z_{\tilde{L}}^{(I+3)i*}Z_N^{1j*}-\frac{m_{l^I}}{\cos\beta m_W}Z_{\tilde{L}}^{Ii*}Z_N^{3j*}.
\end{eqnarray}
The matrices $Z_{\tilde{L}},Z_N$ respectively diagonalize the mass matrices of scalar lepton and neutralino.
The concrete forms of the functions $\mathcal{B}(x,y),\;\mathcal{B}_1(x,y)$ are
\begin{eqnarray}
\mathcal{B}(x,y)=\frac{1}{16 \pi
   ^2}\Big(\frac{x \ln x}{y-x}+\frac{y \ln
   y}{x-y}\Big),~~~
\mathcal{B}_1(x,y)=(
\frac{\partial }{\partial y}+\frac{y}{2}\frac{\partial^2 }{\partial y^2})\mathcal{B}(x,y).
\end{eqnarray}

In a similar way, the corrections from chargino and scalar neutrino are also obtained.
\begin{eqnarray}
&&a_{1}^{\tilde{\nu}\chi^{\pm}}=\sum_{J=1}^3\sum_{i,j=1}^2
\frac{e^2}{s_W^2}\Big[\sqrt{2}\frac{m_{l^I}}{m_W}\texttt{Re}[Z_+^{1j}Z_-^{2j}]|Z_{\tilde{\nu}^{IJ}}^{1i}|^2\sqrt{x_{\chi_j^{\pm}}x_{l^I}}
\mathcal{B}_1(x_{\tilde{\nu}^{Ji}},x_{\chi_j^{\pm}})
\nonumber\\&&\hspace{1.4cm}+\frac{1}{3}(|Z_+^{1j}Z_{\tilde{\nu}^{IJ}}^{1i*}|^2+\frac{m_{l^I}^2}{2m^2_W}|Z_-^{2j*}Z_{\tilde{\nu}^{IJ}}^{1i*}|^2)
x_{\chi_j^{\pm}}x_{l^I}\frac{\partial\mathcal{B}_1(x_{\tilde{\nu}^{Ji}},x_{\chi_j^{\pm}})}{\partial x_{\chi_j^{\pm}}}\Big].
\end{eqnarray}
Here, $Z_-,Z_+$ are used to diagonalize the chargino mass matrix. Because the right-hand neutrino are introduced in BLMSSM, their super partners lead to
six scalar neutrinos. The mass squared matrix of scalar neutrino are diagonalized by $Z_{\tilde{\nu}^{IJ}}$.

Though the Higgs contributions to muon MDM are suppressed by the factor $\frac{m_{l^I}^2}{m^2_W}$, we show their results here.
The one loop Higgs contributions to muon MDM are small. Firstly, we show the analytic results from the neutral Higgs.
\begin{eqnarray}
&&a_{1}^{H^0l}=-\frac{e^2m_{l^I}^2}{2s_W^2m_W^2}\Big[\cos^2\alpha
\Big(x_{l^I}\mathcal{B}_1(x_{H^0},x_{l^I})
-\frac{1}{3}
x_{l^I}^2\frac{\partial}{\partial x_{l^I}}\mathcal{B}_1(x_{H^0},x_{l^I})\Big)\nonumber\\&&
\hspace{1.6cm}+\sin^2\alpha
\Big(x_{l^I}\mathcal{B}_1(x_{h^0},x_{l^I})
-\frac{1}{3}
x_{l^I}^2\frac{\partial}{\partial x_{l^I}}\mathcal{B}_1(x_{h^0},x_{l^I})\Big)
\nonumber\\&&\hspace{1.6cm}
+\cos^2\beta\Big(
x_{l^I}\mathcal{B}_1(x_{G^0},x_{F_1})+\frac{1}{3}x_{l^I}^2\frac{\partial}{\partial x_{l^I}}\mathcal{B}_1(x_{G^0},x_{l^I})\Big)\nonumber\\&&
\hspace{1.6cm}+\sin^2\beta\Big(
x_{l^I}\mathcal{B}_1(x_{A^0},x_{F_1})+\frac{1}{3}x_{l^I}^2\frac{\partial}{\partial x_{l^I}}\mathcal{B}_1(x_{A^0},x_{l^I})\Big)\Big]
\end{eqnarray}
The charged Higgs contributions are written as
\begin{eqnarray}
&&a_{1}^{H^{\pm}\nu}=\sum_{J=1}^3\sum_{i=1}^2\Big[-\frac{1}{3}(|Y_{\nu^J}\cos\beta W_{\nu^{IJ}}^{2i*}|^2+|Y^*_{l^I}\sin\beta U_{\nu^{IJ}}^{1i*}|^2)
x_{l^I}x_{H^\pm}\frac{\partial\mathcal{B}_1(x_{\nu^{Ji}},x_{H^\pm})}{\partial x_{H^\pm}}\nonumber\\&&+\frac{\sin2\beta}{2} \texttt{Re}[Y_{\nu^J}Y_{l^I} W_{\nu^{IJ}}^{2i*} U_{\nu^{IJ}}^{1i}]\sqrt{x_{\nu^{Ji}}x_{l^I}}\Big(x_{G^\pm}\frac{\partial^2 \mathcal{B}(x_{\nu^{Ji}},x_{G^\pm})}{\partial x_{G^{\pm}}^2}
-x_{H^\pm}\frac{\partial^2 \mathcal{B}(x_{\nu^{Ji}},x_{H^\pm})}{\partial x_{H^{\pm}}^2}\Big)\nonumber\\&&
-\frac{1}{3}(|Y_{\nu^J}\sin\beta W_{\nu^{IJ}}^{2i*}|^2+|Y^*_{l^I}\cos\beta U_{\nu^{IJ}}^{1i*}|^2)x_{l^I}x_{G^\pm}\frac{\partial}{\partial x_{G^\pm}}\mathcal{B}_1(x_{\nu^{Ji}},x_{G^\pm})\Big].
\end{eqnarray}
Because of the right handed neutrinos,
the mass matrix of neutrino are expended to $6\times6$. While, the squared mass matrix of scalar
neutrinos turns to $6\times6$ too. The right handed neutrino contributions are very small $(10^{-15}\sim10^{-13})$ and can be neglected
safely.
\subsection{the two-loop Barr-Zee type diagram with a closed scalar loop}
The two-loop Barr-Zee type diagrams can give important contributions to
muon MDM. For the exotic scalar neutrino loops, their effects are suppressed by the Higgs-lepton-lepton
  coupling($\frac{m_{\mu}}{m_W}\sim \frac{1}{1000}$), but are enhanced by the exotic
  scalar neutrino-higgs-exotic scalar lepton coupling including $A_{N_{4}},A_{N_{5}},A_{E_{4}}, A_{E_{5}}$. The concrete expressions
  can be found in Eqs.(30-35). One can find the detailed discussion of ultraviolet properties for these type diagrams' contributions to muon MDM in Ref.\cite{EDM}.
\begin{figure}[h]
\setlength{\unitlength}{1mm}
\centering
\includegraphics[width=4.0in]{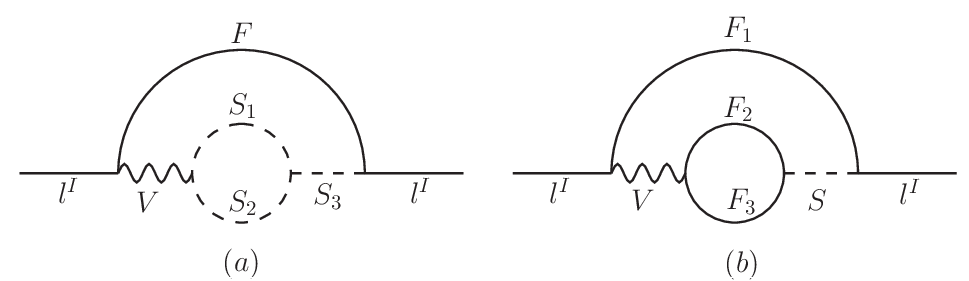}
\caption[]{The two loop Barr-Zee type diagrams with sub Fermion loop
and sub scalar loop.}
\end{figure}

\subsubsection{The virtual particles in the scalar loop are all neutral particles}
 The closed scalar virtual particles are neutral Higgs, scalar neutrinos and exotic scalar neutrinos, and they are attached to
 Z (neutral Higgs). With CP-even Higgs between the scalar loop and Fermion line, the two loop diagrams Fig.2(a) with scalar neutrinos (exotic scalar neutrinos) give contributions to lepton MDM as
  \begin{eqnarray}
&&a^{\tilde{\nu},H^0}_2(Z)=- \frac{e^3(1-4s_W^2)x_{l^I}}{4s_W^3c_W^2\sqrt{x_W}}
\sum_{S=\tilde{\nu},\tilde{N}'_{4,5}}\sum_{i,j,k=1}^2 \texttt{Re}[(Z_{S}^\dag)^{i1}Z_{S}^{1j}\frac{H_{H^0_kS_iS_j}}{M_{NP}}Z_R^{1k}]\nonumber\\&&\hspace{1.8cm}\times \Big(4\mathcal{P}_1
+x_{l^I}\mathcal{P}_2\Big)(x_Z,x_{H^0_k}
,x_{l^I},x_{S_i},x_{S_{j}}),
\end{eqnarray}
The couplings $H_{H^0_kS_iS_j}$ between CP-even Higgs and exotic scalar neutrinos$(\tilde{N}'_{4,5})$, can be found in
Eqs.(\ref{LHLpLp}), (\ref{LHLpLp1}). Because the super fields $\tilde{N}^c$ is introduced in BLMSSM, the couplings related with the MSSM scalar
neutrinos are changed, and they are
corrected as
\begin{eqnarray}
&&H_{H^0_1\tilde{\nu}_i\tilde{\nu}_j}=(N^u_M)_{ij}\sin\alpha+(N^d_M)_{ij}\cos\alpha;~~~~~
H_{H^0_2\tilde{\nu}_i\tilde{\nu}_j}=(N^u_M)_{ij}\cos\alpha-(N^d_M)_{ij}\sin\alpha,\nonumber\\&&
(N^u_M)_{ij}=V_{EW}\sin\beta\Big(\frac{e^2}{4s_W^2c_W^2}(Z_{\tilde{\nu}}^\dag)^{i1}Z_{\tilde{\nu}}^{1j}
-|Y_{\nu}|^2\delta_{ij}\Big)+\lambda_{\nu_c}^*\bar{v}_L
  (Z_{\tilde{\nu}}^\dag)^{i2}Z_{\tilde{\nu}}^{1j}-\frac{A_N}{\sqrt{2}}(Z_{\tilde{\nu}}^\dag)^{i2}Z_{\tilde{\nu}}^{1j},\nonumber\\&&
  (N^d_M)_{ij}=-\frac{e^2}{4s_W^2c_W^2}V_{EW}\cos\beta (Z_{\tilde{\nu}}^\dag)^{i1}Z_{\tilde{\nu}}^{1j}-\frac{\mu^*}{\sqrt{2}}Y_{\nu}.
  \end{eqnarray}
The concrete forms of the functions $\mathcal{P}_1,\mathcal{P}_2$ are
  \begin{eqnarray}
&&\mathcal{P}_1(v,t,f,s,w)= \frac{ f (s-w)}{8}\frac{\partial^2}{\partial f^2}\Big(
 \mathcal{B}(s,w)
  \mathcal{ C}(v,t,f)\nonumber+\mathcal{F}(v,t,f,s,w
   )+\frac{\mathcal{ C}_1(v,t,f)}{
   16 \pi ^2}\Big),\nonumber\\&&
  \mathcal{P}_2(v,t,f,s,w) =-\frac{4}{3}(2
    +f\frac{\partial}{\partial f})\frac{1}{f}\mathcal{P}_1(v,t,f,s,w).
 \end{eqnarray}
The functions $\mathcal{C},\mathcal{C}_1,\mathcal{F}$ are collected in the appendix.

In this type, when the scalar particles between the scalar loop and Fermion line are CP-odd Higgs,
the contributions from scalar neutrinos (exotic scalar neutrinos) read as
 \begin{eqnarray}
&&a_2^{\tilde{\nu},A^0}(Z)=-\frac{e^3x_{l^I}}{4s_W^3c_W^2\sqrt{x_W}}\sum_{S=\tilde{\nu},\tilde{N}'}\sum_{i,j,k=1}^2 \texttt{Re}[Z_H^{1k}(Z_{S}^\dag)^{i1}Z_{S}^{1j}\frac{
H_{A^0_kS_iS_j}}{M_{NP}}]\nonumber\\&&\times\Big(4\mathcal{P}_1
-x_{l^I}\mathcal{P}_2
\Big)(x_Z,x_{A^0_k}
,x_{l^I},x_{S_i},x_{S_j}).
\end{eqnarray}
One can find CP-odd Higgs and exotic scalar neutrinos $(\tilde{N}')$ couplings $H_{A^0_kS_iS_j}$ in
Eq.(\ref{LALpLp}). The concrete forms for the couplings between CP-odd Higgs and MSSM scalar neutrinos are corrected as
\begin{eqnarray}
&&H_{A^0_1\tilde{\nu}_i\tilde{\nu}_j}=\cos\beta (P^u_N)_{ij}-\sin\beta(P^d_N)_{ij},~~~
(P^d_N)_{ij}=-\frac{\mu^*}{\sqrt{2}}Y_{\nu}(Z_{\tilde{\nu}}^\dag)^{i2}Z_{\tilde{\nu}}^{1j},\nonumber\\
&&H_{A^0_2\tilde{\nu}_i\tilde{\nu}_j}=\sin\beta (P^u_N)_{ij}+\cos\beta(P^d_N)_{ij} ,~~~ (P^u_N)_{ij}=-(\lambda_{\nu_c}^*\bar{v}_L-\frac{A_N}{\sqrt{2}})(Z_{\tilde{\nu}}^\dag)^{i2}Z_{\tilde{\nu}}^{1j}.
\end{eqnarray}
When the scalar particles in the scalar loop are all neutral Higgs, the corrections to lepton MDM from
Fig.2(a) are
\begin{eqnarray}
&&a^{H^0,A^0}_2(Z)=\frac{-e^5 x_{l^I}}{16s_W^5c_W^4\sqrt{x_{W}}}\sum_{i,j,k=1}^2
\texttt{Re}[\frac{B^{i}_{R}}{M_{{NP}}}A^{ij}_{M}A^{jk}_{H}Z^{1k}_{H}]\nonumber\\&&\times\Big(4\mathcal{P}_1
- x_{l^I}\mathcal{P}_2
\Big)(x_Z,x_{A_k^0}
,x_{l^I},x_{H_i^0},x_{A_j^0}),
\end{eqnarray}
where the couplings $B^{i}_{R},A^{ij}_{M},A^{jk}_{H}$  can be found in Ref.\cite{MSSM}.

\subsubsection{The vector is $\gamma$ (Z), and the scalar loop are charged scalar particles}
When the vector is photon, contributions from the Fig.2(a) are just produced by the neutral CP even Higgs.
That is to say the corresponding CP odd Higgs' contribution is zero.
\begin{eqnarray}
&&a^{S,H^0}_{2}(\gamma)
=-\frac{8Q_Se^3 x_{l^I} }{s_W\sqrt{x_W}}\sum_{i=1}^2\Big(\sum_{S=\tilde{L},\tilde{U},\tilde{D},\tilde{L}',\tilde{\mathcal{U}},\tilde{\mathcal{D}},
H^{\pm}}\Big)\texttt{Re}[Z_R^{1i}\frac{H_{H^0_i SS}}{M_{NP}}]
\nonumber\\&&\times\Big((Q_S\mathcal{W}_{1}+\mathcal{P}_1)
+x_{l^I}(Q_S\mathcal{W}_{2}+\frac{1}{4}\mathcal{P}_2)\Big)(0,x_{l^I},x_{H^0_i},x_S,x_S).
\end{eqnarray}
Here, $Q_S$ is the electric charge of the scalar particles. One can find the functions $\mathcal{W}_{1}, \mathcal{W}_{2}$ in
the appendix. $H_{H^0_i SS}$ are the couplings for CP even Higgs and two scalar particles. With S representing the MSSM
particles $\tilde{L},\tilde{U}, \tilde{D},H^{\pm}$, the concrete forms of $H_{H^0_i SS}$ are in Ref.\cite{MSSM}. We have deduced the
coupling between Higgs and exotic scalar quarks in our previous work\cite{weBLMSSM}. The couplings for charged exotic scalar leptons and
neutral Higgs are shown in Eq.(\ref{LHLpLp}).
The exotic scalar quark loop contributions are also suppressed by the factor $\frac{m_{\mu}}{m_W}\sim \frac{1}{1000}$
, and they are increased by the coupling of Higgs-exotic scalar quarks-exotic scalar quarks which includes $A_{d_4},A_{d_5},A_{u_4},A_{u_5}$ et al.
  Their concrete forms can be found in Eqs.(36-38).

For the Fig.2(a), with vector $Z$, both CP even and CP odd Higgs give corrections to the lepton MDM, and their results are obtained here.
\begin{eqnarray}
&&a^{S,H^0}_{2}(Z)=\frac{e^3x_{l^I}}{s_W^3c_W^2\sqrt{x_W}}\sum_{i=1}^2\Big(\sum_{S_1,S_2=\tilde{L},\tilde{U},\tilde{D},\tilde{L}',\tilde{\mathcal{U}},\tilde{\mathcal{D}},
H^{\pm}}\Big)\Big[
\texttt{Re}[\frac{H_{H^0 S_1S_2}}{M_{NP}}G_{Z S_1S_2}Z_R^{1i}]\nonumber\\&&\times(1-4s_W^2)\Big((Q_S\mathcal{W}_{1}
+\mathcal{P}_1)+x_{l^I}(Q_S\mathcal{W}_{2}+\frac{\mathcal{P}_2}{4})\Big)(x_Z,x_{l^I},x_{H^0_i},x_{S_1},x_{S_2})
\nonumber\\
&&+\texttt{Re}[\frac{H_{A^0 S_1S_2}}{M_{NP}}G_{Z S_1S_2}Z_H^{1i}]\Big((Q_S\mathcal{W}_{1}+\mathcal{P}_1)
-x_{l^I}(Q_S\mathcal{W}_{2}+\frac{\mathcal{P}_2 }{4})\Big)(x_Z,x_{l^I},x_{A^0_i},x_{S_1},x_{S_2})\Big],
\end{eqnarray}
$H_{A^0 S_1S_2}$( $G_{Z S_1S_2}$) are the couplings for two scalar particles and CP odd Higgs ($Z$). When $S_1,S_2$ are
MSSM scalar particles, $H_{A^0 S_1S_2}$( $G_{Z S_1S_2}$) can be found in Ref.\cite{MSSM}.
The scalar exotic charged lepton and CP-odd
Higgs (Z) coupling are in Eqs.(\ref{LALpLp}), (\ref{VSSLPLP}). Eq.(\ref{VSSQpQP}) gives out Z and two exotic scalar quarks coupling.
  \subsubsection{The vector is $W^{\pm}$, and the scalar loop has charged particles}
 Charged scalar lepton and scalar quark in MSSM contribute to lepton MDM. In the same way, we also obtain the exotic scalar lepton and exotic
 scalar quark contributions.
 \begin{eqnarray}
&&a_{2}^{S,H^{\pm}}(W^{\pm})=-\frac{4e^2}{s_W^2}\sum_{\alpha, i=1}^2\sum_{S_{1}=\tilde{L},
\tilde{L}',\tilde{D},\tilde{\mathcal{D}},H^{\pm}}\sum_{S_{2}
=\tilde{\nu},\tilde{N}',\tilde{U},\tilde{\mathcal{\mathcal{U}}},H^0,A^0}
\Big(\sqrt{x_{l^I}}\texttt{Re}[G_{WS_1S_2}(U^{\dag}_{\nu^{J}})^{\alpha 1}\nonumber\\&&\times\frac{H_{H_i^{\pm}S_1S_2}}{M_{NP}}A_{H_i^{\pm}\nu l}\mathcal{W}_{3}
]\Big)(x_{\nu},x_{W},x_{H^{\pm}_i},x_{S_1},x_{S_2},Q_{S_1},Q_{S_2}).
\end{eqnarray}
The complex functions $\mathcal{W}_3,\mathcal{W}_4$ are collected in the appendix. The concrete forms for the couplings
related with neutrino and scalar neutrino are different from those in MSSM.
\begin{eqnarray}
&&
A_{H^{\pm}\nu l}=\sin\beta Y_l^I U_{\nu^{IJ}}^{1i},~~B_{H^{\pm}\nu l}=\cos\beta Y_{\nu}^*W_{\nu^{IJ}}^{2i},~~A_{G^{\pm}\nu l}=\cos\beta Y_l^I U_{\nu^{IJ}}^{1i},\nonumber\\&&
B_{G^{\pm}\nu l}=\sin\beta Y_{\nu}^*W_{\nu^{IJ}}^{2i},~~H_{H^{\pm}\tilde{L}\tilde{\nu}}=(L_M^u)_{ij}\cos\beta-(L_M^d)_{ij}\sin\beta,\nonumber\\&&
H_{G^{\pm}\tilde{L}\tilde{\nu}}=(L_M^u)_{ij}\sin\beta+(L_M^d)_{ij}
\cos\beta,~~~G_{W\tilde{L}\tilde{\nu}}=(Z_{\nu^{IJ}}^{\dag})^{i1}Z_L^{IJ},
\end{eqnarray}
with
\begin{eqnarray}
  &&(L^u_M)_{ij}=(-\frac{e^2}{2\sqrt{2}s_W^2}+|Y_{\nu}|^2)V_{EW}\sin\beta (Z_{\tilde{\nu}}^\dag)^{i1}Z_{\tilde{L}}^{1j}
  -Y_e^*\mu (Z_{\tilde{\nu}}^\dag)^{i1}Z_{\tilde{L}}^{2j}\nonumber\\&&
  +\frac{V_{EW}\cos\beta}{\sqrt{2}}Y_e^*Y_{\nu} (Z_{\tilde{\nu}}^\dag)^{i2}Z_{\tilde{L}}^{2j}+
  (A_N-\sqrt{2}\lambda_{\nu_c}^*\bar{v}_L)(Z_{\tilde{\nu}}^\dag)^{i2}Z_{\tilde{L}}^{1j},\nonumber\\&&
  (L^d_M)_{ij}=(-\frac{e^2}{2\sqrt{2}s_W^2}+|Y_{e}|^2)V_{EW}\cos\beta (Z_{\tilde{\nu}}^\dag)^{i1}Z_{\tilde{L}}^{1j}
  -Y_{\nu}\mu^* (Z_{\tilde{\nu}}^\dag)^{i2}Z_{\tilde{L}}^{1j}\nonumber\\&&+\frac{V_{EW}\sin\beta}{\sqrt{2}}Y_e^*Y_{\nu} (Z_{\tilde{\nu}}^\dag)^{i2}Z_{\tilde{L}}^{2j}+
  A_E^*(Z_{\tilde{\nu}}^\dag)^{i1}Z_{\tilde{L}}^{2j},
  \end{eqnarray}
The other necessary concrete forms for the couplings  $G_{WS_1S_2}, H_{H_i^{\pm}S_1S_2}$ can be found in Eqs.(\ref{VSSLPLP}),(\ref{VSSQpQP}),(\ref{HPMSSLP}),(\ref{HPMSSLP1}),(\ref{HPMSSQP}),(\ref{HPMSSQP1}) and Ref.\cite{weBLMSSM}.

\subsection{the two-loop Barr-Zee type diagram with a closed Fermion loop}
When the inserted is a  fermion loop, the diagrams can be divided into two parts, according to the
vector neutral Boson$(\gamma, Z)$ and charged one$(W^\pm)$. For $H^{\pm}F_1F_2$ coupling, it becomes large with the
heavy Fermion mass, and may give important contributions.
\subsubsection{the vector is $\gamma, Z$, and the Fermion loop are all charged particles}
When the fermion loop is quarks and exotic quarks, charged leptons and exotic charged leptons,
the contributions for muon MDM from the two loop diagrams with charged fermion loop inserted between $\gamma$
and CP even Higgs are
 \begin{eqnarray}
 &&a_2^{F,H^0}(\gamma)
=\frac{16e^3x_{l^I}}{s_W \sqrt{x_{W}}}\sum_{i=1}^2\sum_{F=b,t,\chi^{\pm},\tau,L',b',t'}Q_F^2Z_R^{1i} \texttt{Re}[Y_{H^0_iFF}]
\nonumber\\&&\times\sqrt{x_{F}}(\mathcal{W}_5+ \mathcal{W}_6
+2x_{l^I}\mathcal{W}_7)(0,x_{l^I},x_{H^0_i},x_{F},x_{F}),
 \end{eqnarray}
 where $Y_{H^0_iFF}$ are the right hand parts of the couplings between the CP-even Higgs and the Fermions$(b,t,\chi^{\pm},\tau,L',b',t')$, and
 the general form is written as $i(J_{H^0_iFF}\omega_-+Y_{H^0_iFF}\omega_+)$. The concrete forms of $Y_{H^0_iFF}$ with $F=(b,t,\chi^{\pm},\tau,L',b',t')$
 can be found in Ref.\cite{MSSM,weBLMSSM}. $Y_{H^0_iFF}$ for $F=L'$ are shown in Eq.(\ref{HLL}). To save space in the text,
 the form factors$\mathcal{W}_{5,6\dots 19}$ are shown in the appendix.

 In the same way, we get the two loop contributions with $\gamma$, CP odd Higgs and charged fermion loop.
  \begin{eqnarray}
  &&a_2^{F,A^0}(\gamma)
=\frac{16e^3x_{l^I}}{s_W \sqrt{x_{W}}}\sum_{i=1}\sum_{F=b,t,\chi^{\pm},\tau,L',b',t'}^2Q_F^2 \texttt{Re}[Y_{A^0_iFF}Z_H^{1i}]
\nonumber\\&&\times\sqrt{x_{F}}(\mathcal{W}_8+\mathcal{W}_9-x_{l^I}
\mathcal{W}_{10})(0,x_{l^I},x_{A^0_i},x_{F},x_{F}).
 \end{eqnarray}
   $Y_{A^0_iFF}$ are the right hand parts of the couplings between the CP-odd Higgs
  with $F=(b,t,\chi^{\pm},\tau,b',t')$, whose forms are obtained in the same way as  $Y_{H^0_iFF}$.

When the vector is Z instead of $\gamma$, the corresponding expressions of the MDM contribution are more complex.
The results from the CP-even Higgs, Z and charged fermion loop at two loop level are
\begin{eqnarray}
 &&a_2^{F,H^0}(Z)
=2\frac{-Q_Fe^3x_{l^I}}{s_W^2c_W \sqrt{x_{W}}}\sum_{i=1}^2\sum_{F=b,t,\chi^{\pm},\tau,L',b',t'}Z_R^{1i}\Big\{\texttt{Re}[(H_{ZF_1F_2} Y_{H_i^0F_1F_2}\nonumber\\&&-J_{H_i^0F_1F_2} T_{ZF_1F_2})]\sqrt{x_{F_1}}\Big(\mathcal{W}_8
+x_{l^I}\frac{1}{2}\mathcal{W}_{10}\Big)(x_{Z},x_{l^I},x_{H^0_i},x_{F_1},x_{F_2})\nonumber\\&&
+\texttt{Re}[( T_{ZF_1F_2} Y_{H_i^0F_1F_2}-H_{ZF_1F_2} J_{H_i^0F_1F_2}
   )]\sqrt{x_{F_2}}\Big(\mathcal{W}_9(x_{Z},x_{l^I},x_{H^0_i},x_{F_1},x_{F_2})\nonumber\\&&
   +x_{l^I}\frac{1}{2}\mathcal{W}_{10}(x_{Z},x_{l^I},x_{H^0_i},x_{F_2},x_{F_1})\Big)+
\texttt{Re}[(1-4s_W^2)(H_{ZF_1F_2} J_{H_i^0F_1F_2} \nonumber\\&&+ T_{ZF_1F_2} Y_{H_i^0F_1F_2})]\sqrt{x_{F_2}}\Big(\mathcal{W}_5
+x_{l^I}\mathcal{W}_7\Big)(x_{Z},x_{l^I},x_{H^0_i},x_{F_1},x_{F_2})\nonumber\\&&
+\texttt{Re}[(1-4s_W^2)(J_{H_i^0F_1F_2} T_{ZF_1F_2} + H_{ZF_1F_2} Y_{H_i^0F_1F_2})]\sqrt{x_{F_1}}
\nonumber\\&&\times\Big(\mathcal{W}_6(x_{Z},x_{l^I},x_{H^0_i},x_{F_1},x_{F_2})+x_{l^I}\mathcal{W}_7(x_{Z},x_{l^I},x_{H^0_i},x_{F_2},x_{F_1}))
\Big\}.
 \end{eqnarray}
 Generally, $ZF_1F_2$ couplings are expressed as $ie(H_{ZF_1F_2}\gamma_{\alpha}\omega_-+T_{ZF_1F_2}\gamma_{\alpha}\omega_+)$.
 One can obtain $H_{ZF_1F_2},T_{ZF_1F_2}$ for $F=(L',b',t')$ in Eqs.(\ref{VLL},\ref{VQQ}).  $Y_{H_i^0F_1F_2}$
are similar with $Y_{H_i^0FF}$, while $J_{H_i^0F_1F_2}$ are couplings of the left parts.

We also obtain the CP-odd Higgs contribution from the diagram with vector Z and charged fermion loop shown in Fig.2(b).
 \begin{eqnarray}
 &&a_2^{F,A^0}(Z)
=2\frac{-Q_Fe^3x_{l^I}}{s_W^2c_W \sqrt{x_{W}}}\sum_{i=1}^2\sum_{F=b,t,\chi^{\pm},\tau,L',b',t'}Z_H^{1i}
\Big\{(1-4s_W^2)\texttt{Re}[( T_{ZF_1F_2} Y_{A_i^0F_1F_2}\nonumber\\&&-H_{ZF_1F_2} J_{A_i^0F_1F_2}
   )]\sqrt{x_{F_2}}\Big(\mathcal{W}_9(x_{Z},x_{l^I},x_{A^0_i},x_{F_1},x_{F_2})
 -\frac{x_{l^I}}{2}\mathcal{W}_{10}(x_{Z},x_{l^I},x_{A^0_i},x_{F_2},x_{F_1})\Big)
\nonumber\\&&+\texttt{Re}[(H_{ZF_1F_2} J_{A_i^0F_1F_2} + T_{ZF_1F_2} Y_{A_i^0F_1F_2})]\sqrt{x_{F_2}}\Big(\mathcal{W}_5-x_{l^I}\mathcal{W}_7\Big)(x_{Z},x_{l^I},x_{A^0_i},x_{F_1},x_{F_2})\nonumber\\&&
+\texttt{Re}[(J_{A_i^0F_1F_2} T_{ZF_1F_2} + H_{ZF_1F_2} Y_{A_i^0F_1F_2})]\sqrt{x_{F_1}}
\Big(\mathcal{W}_6(x_{Z},x_{l^I},x_{A^0_i},x_{F_1},x_{F_2})\nonumber\\&&
-x_{l^I}\mathcal{W}_7(x_{Z},x_{l^I},x_{A^0_i},x_{F_2},x_{F_1})\Big) +(1-4s_W^2)\texttt{Re}[(H_{ZF_1F_2} Y_{A_i^0F_1F_2}\nonumber\\&&-J_{A_i^0F_1F_2} T_{ZF_1F_2})]
\sqrt{x_{F_1}}\Big(\mathcal{W}_8-x_{l^I}\frac{1}{2}\mathcal{W}_{10}\Big)(x_{Z},x_{l^I},x_{A^0_i},x_{F_1},x_{F_2})
\Big\}.
 \end{eqnarray}
 Similarly, we get the CP odd Higgs couplings $Y_{A_i^0F_1F_2},J_{A_i^0F_1F_2}$.
\subsection{the vector is $W^{\pm}$, and the fermion loop have charged particles}
 The contribution to lepton MDM from the diagram with vector $W^{\pm}$ and Fermion loop
 are obtained here.
\begin{eqnarray}
&&a_{2}^{F,H^{\pm}}(W^{\pm})
=\sum_{i=1}^2\sum_{F_1=b,b',\tau,L',\chi^{\pm}}\sum_{F_2=t,t',\nu_{\tau},N',\chi^0}\frac{4e^2}{s_W^2}\Big(
\texttt{Re}[A_{H_i^{\pm}\nu l}(U^{\dag}_{\nu^{J}})^{\alpha 1}(T_{WF_1F_2}Y_{H_i^{\pm}F_1F_2}\mathcal{W}_{12}\nonumber\\&&
+H_{WF_1F_2}
J_{H_i^{\pm}F_1F_2}\mathcal{W}_{13})]\sqrt{x_{l^I} x_{F_2}}
+\sqrt{x_{l^I} x_{F_1}}\texttt{Re}[A_{H_i^{\pm}\nu l}(U^{\dag}_{\nu^{J}})^{\alpha 1}(H_{WF_1F_2}Y_{H_i^{\pm}F_1F_2}\mathcal{W}_{14}\nonumber\\&&+J_{H_i^{\pm}F_1F_2}T_{WF_1F_2}\mathcal{W}_{15})]
\Big)(x_W,x_\nu,x_{H^{\pm}_i},x_{F_1},x_{F_2},Q_{F_1},Q_{F_2}).
\end{eqnarray}
The couplings related with exotic leptons (quarks) are given in Eq.(\ref{HpmLL}), (\ref{VLL}), (\ref{VQQ}) and Ref.\cite{smneutron}.

Because the right handed neutrino is introduced in BLMSSM, the couplings related with neutrino are
not same as those in MSSM. We deduced the needed couplings here.
\begin{eqnarray}
&&H_{W\tau\nu_{\tau}}=(U_{\nu^{IJ}}^\dag)^{i1},~ F_{G^{\pm}L\nu}=-Y_{\nu}\sin\beta (W_{\nu^{IJ}}^\dag)^{i2},~~~Y_{G^{\pm}L\nu}=Y_l^{I*}\cos\beta (U_{\nu^{IJ}}^\dag)^{i1},\nonumber\\&&T_{W\tau\nu_{\tau}}=0,~~~~~
F_{H^{\pm}L\nu}=-Y_{\nu}\cos\beta (W_{\nu^{IJ}}^\dag)^{i2},~~~Y_{H^{\pm}L\nu}=-Y_l^{I*}\sin\beta (U_{\nu^{IJ}}^\dag)^{i1}.
\end{eqnarray}

\section{the numerical results}
In this section, we show our numerical results. For the input
parameters, we take into account the experimental constraints
from the  lightest neutral CP even Higgs $m_{_{h^0}}\simeq125.7\;{\rm GeV}$
and the neutrino experiment data:
\begin{eqnarray}
&&\sin^22\theta_{13}=0.090\pm 0.009,~~\sin^2\theta_{12} =0.306_{-0.015}^{+0.018},
~~\sin^2\theta_{23}=0.42_{-0.03}^{+0.08},\nonumber\\
&&\Delta m_{\odot}^2 =7.58_{-0.26}^{+0.22}\times 10^{-5} {\rm eV}^2,
~~|\Delta m_{A}^2| =2.35_{-0.09}^{+0.12}\times 10^{-3} {\rm eV}^2.
\label{neu-oscillations2}
\end{eqnarray}

In our previous work, we fit the neutrino experiment data shown as Eq.(\ref{neu-oscillations2}) in BLMSSM \cite{BiaoChen}.
The lepton flavor violation is taken into account through $(Y_{\nu})_{ij}, (m_{\tilde{\nu}^c}^2)_{ij},(m_L^2)_{ij},(m_R^2)_{ij}, (i,j=1,2,3)$. In the numerical
discussion, the non-diagonal elements of these matrixes are not zero.
Therefore, the lepton flavor violation are considered and there is a transition between muon-sneutrinos and tau-sneutrinos.

Firstly, we give out the SM relevant parameters.
\begin{eqnarray}
&&\alpha_s(m_Z)=0.118,\;\alpha(m_Z)=1/128,
\;s_W^2(m_Z)=0.23,\;m_W=80.4{\rm GeV},\nonumber\\
&&m_Z=91.2{\rm GeV},\;
m_t=174.2{\rm GeV},\;m_b=4.2{\rm GeV},\;
m_u=2.3\times10^{-3}{\rm GeV},\;\;\nonumber\\&&m_d=4.8\times10^{-3}{\rm GeV},\;\;
m_s=0.95\times10^{-3}{\rm GeV},\;\; m_c=1.275{\rm GeV},\nonumber\\&&
m_e=0.51\times10^{-3}{\rm GeV},\;\;m_{\mu}=0.105{\rm GeV},\;\;m_{\tau}=1.777{\rm GeV},
\nonumber\\&&
{\rm CKM}_{11}=0.9743,\;\;{\rm CKM}_{22}=0.9734,\;\;{\rm CKM}_{33}=0.9991.
\label{PDG-SM}
\end{eqnarray}
Because the muon MDM is related with the real parts of the results, to simplify the numerical discussion we suppose
all the involved parameters in BLMSSM are real.

 The used parameters in BLMSSM are collected here.
\begin{eqnarray}
&&\tan\beta{_B}=\tan\beta_{L}=2,\;~~B_4=L_4={3\over2},\;\tan\beta=15,
\nonumber\\
&&m_{\tilde{Q}_3}=m_{\tilde{U}_3}=m_{\tilde{D}_3}=1.4{\rm TeV},\;~~m_{Z_B}=m_{Z_L}=1{\rm TeV},
\nonumber\\
&&m_{\tilde{U}_4}=m_{\tilde{D}_4}=m_{\tilde{Q}_5}=m_{\tilde{U}_5}
=m_{\tilde{D}_5}=1{\rm TeV},\;\; m_{\tilde{Q}_4}=790{\rm GeV},\nonumber\\
&&m_{\tilde{L}_4}=m_{\tilde{\nu}_4}=m_{\tilde{E}_4}=m_{\tilde{L}_5}=m_{\tilde{\nu}_5}
=m_{\tilde{E}_5}=1.4{\rm TeV}\;,\nonumber\\
&&A_{u_4}=A_{u_5}=A_{d_4}=A_{d_5}=550{\rm GeV}
\;,A_{b}=A_{t}=-1{\rm TeV},\nonumber\\
&&\upsilon_{B_t}=\sqrt{\upsilon_B^2+\overline{\upsilon}_B^2}=3{\rm TeV}\;,\;\;
\upsilon_{L_t}=\sqrt{\upsilon_{L}^2+\overline{\upsilon}_{L}^2}=3{\rm TeV}\;,
\nonumber\\
&&~m_{1}=1{\rm TeV},\;m_2=750{\rm GeV},\; \mu_H=-800{\rm GeV}.
\nonumber\\
&&Y_{u_4}=0.8Y_t,\;Y_{d_4}=0.7Y_b,\;Y_{u_5}=0.7Y_b,\;Y_{d_5}=0.1Y_t,
\nonumber\\
&&A'_e=A'_{\mu}=A'_{\tau}=130{\rm GeV},\lambda_{\nu^c}=1,\;A_{\nu_4}=A_{\nu_5}=550 {\rm GeV},\nonumber\\&&
A'_u=A'_c=A'_t=A'_d=A'_s=A'_b=500{\rm GeV},\nonumber\\&&
Y_{\nu_4}=0.6,Y_{\nu_5}=1.1,Y_{e_4}=1.3,Y_{e_5}=0.6,\mu_L=500{\rm GeV}\nonumber\\&&
A_{\nu_e}=A_{\nu_\mu}=A_{\nu_\tau}=
A_{\nu_e^c}=A_{\nu_\mu^c}=A_{\nu_\tau^c}=-500{\rm GeV}.
\end{eqnarray}
We suppose the following relations in the numerical discussion, then the numerical discussion
is simplified.
\begin{eqnarray}
&&A_e=A_{\mu}=A_{\tau}=A_L,~~
A_c=A_s=A_{cs},\nonumber\\&&
m_{\tilde{e}}=m_{\tilde{\mu}}=m_{\tilde{\tau}}=
m_{\tilde{\nu}_e}=m_{\tilde{\nu}_\mu}=m_{\tilde{\nu}_\tau}=ML_s
\nonumber\\&&A_{e_4}=A_{e_5}=AE_{45},\; m_{\tilde{Q}_2}=m_{\tilde{U}_2}=m_{\tilde{D}_2}=MQ_2;
\end{eqnarray}

In order to reflect the flavor mixing obviously and simplify the discussion, we define the off-diagonal elements in the following form.
\begin{eqnarray}
&&(m_{\tilde{\nu}^c}^2)_{ij}=(m_L^2)_{ij}=(m_R^2)_{ij}=MLa^2,
~(AL)_{ij}=ALa, \texttt{ with}~ i\neq j,(i,j=1,2,3).
\end{eqnarray}
When $MLa=0, ALa=0$ there is no flavor mixing for the scalar leptons.

\subsection{the one loop numerical results}

Comparing with the two loop contributions from the new physics,
the one loop new physics contributions to muon MDM are dominant.
Therefore, the parameters having relation with the one loop contributions
should affect the results apparently. $ML_s$ is in the squared mass matrixes
 of scalar charged leptons and scalar neutrinos, and these scalar particles
 can give one loop corrections to muon MDM.

\begin{figure}[h]
\setlength{\unitlength}{1mm}
\centering
\includegraphics[width=4.0in]{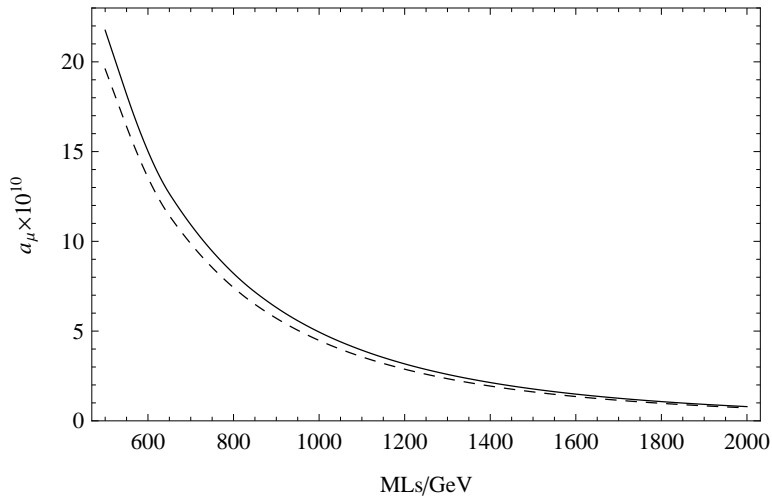}
\caption[]{The one loop scalar lepton and neutralino contributions to muon MDM,
 the dashed-line and solid-line represent muon MDM varying with $ML_s$, for
$A_L=-500{\rm GeV}$ and $A_L=-800{\rm GeV}$ respectively.}\label{oSLML}
\end{figure}

 At first, supposing $MLa=0$ and $ALa=0$, we study the one loop contributions individually with the varying $MLs$(500-2000GeV).
 The one loop scalar lepton-neutralino contributions are plotted in Fig.\ref{oSLML}.

The one loop scalar lepton-neutralino contributions are about $4.5\times10^{-10}$ when $MLs=1000$GeV.
The contributions turn large with the decreasing $MLs$. They can reach $20\times10^{-10}$ with $MLs=500$GeV.
If $MLs$ turns smaller than 500GeV, these contributions can be much larger. As $MLs>1600$ GeV, the results turn
 small, which are about $1.7\times10^{-10}$. The value of $AL$ affects the results slightly. From Fig.\ref{oSLML}, one can find the
 dotted line is up the solid line to small extent.

\begin{figure}[h]
\setlength{\unitlength}{1mm}
\centering
\includegraphics[width=4.0in]{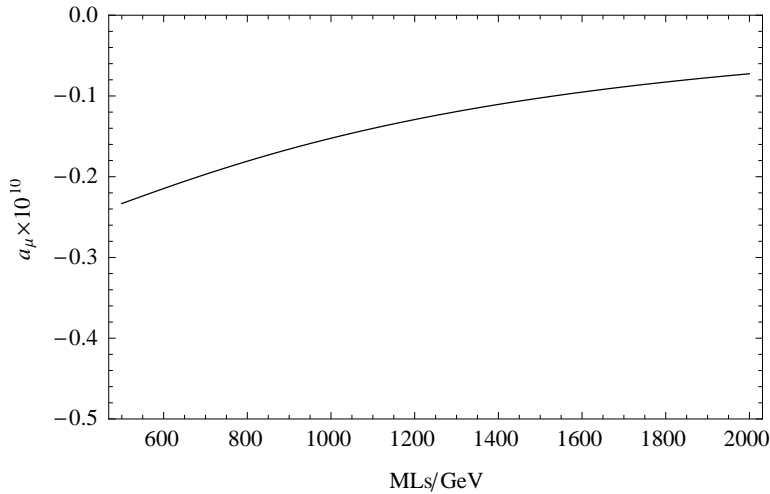}
\caption[]{The one loop scalar neutrino-chargino contributions to muon MDM,
 the solid-line and dashed-line represent muon MDM varying with $ML_s$, for
$A_L=-500{\rm GeV}$ and $A_L=-800{\rm GeV}$ respectively.}\label{oSnuML}
\end{figure}

The one loop scalar neutrino-chargino contributions are plotted in Fig.\ref{oSnuML}.
The dashed line and solid line are coincident. The scalar neutrino-chargino one loop contributions
are about $-2.0\times10^{-11}$ which are approximately one order smaller than the one loop scalar lepton-neutralino contributions.
Obviously, these results vary slightly with $MLs$.

The neutral Higgs-lepton and charged Higgs-neutrino one loop contributions are both at the order of $10^{-14}$. Therefore, one
should not consider them.  Then the  one loop scalar lepton-neutralino contributions are dominant.
In MSSM, the one loop contribution to muon MDM is approximately
\begin{eqnarray}
13\times 10^{-10}\Big(\frac{100\texttt{GeV}}{M_{SUSY}}\Big)^2\tan\beta \texttt{sign}(\mu_H).
\end{eqnarray}
With $M_{SUSY}=1000$ GeV, and $\tan\beta=15$, the MSSM one loop contribution to muon MDM is about
$2.0\times 10^{-10}$. In our results, when $\tan\beta=15$ and $MLs=1000$ GeV, the BLMSSM one loop result is
about $4.5\times10^{-10}$. Roughly speaking, the BLMSSM one loop result accords with MSSM one loop estimate.
Strictly speaking, the BLMSSM one loop result is abut double as MSSM one loop estimate. What is the reason?
When $MLs=1000$ GeV, half of the scalar lepton masses are about 800GeV, the others are about 1000GeV, and the neutralinos masses are obout
700GeV, which enhance the BLMSSM contributions. So using MSSM estimate formula with $M_{SUSY}\sim 700$GeV we can obtain the one loop
contribution $4.0\times 10^{-10}$. On the whole, the BLMSSM one loop results confirm with the one loop MSSM estimate.

\begin{figure}[h]
\setlength{\unitlength}{1mm}
\centering
\includegraphics[width=4.0in]{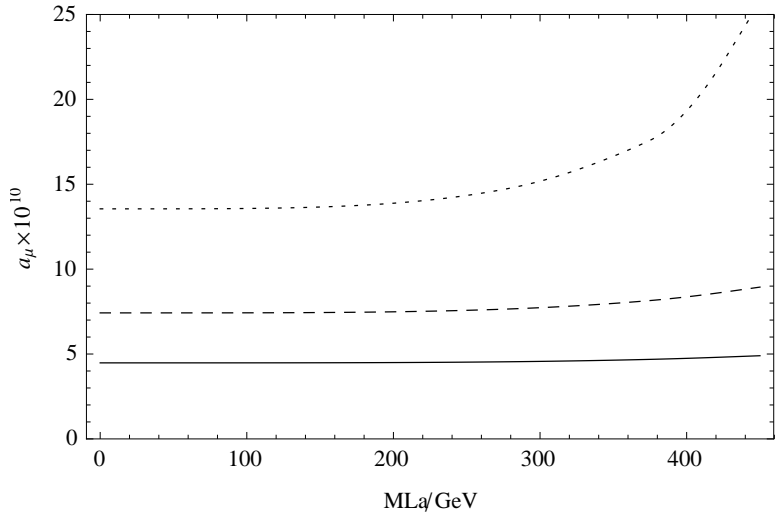}
\caption[]{The one loop scalar lepton-neutralino contributions to muon MDM,
 the dotted-line, dashed-line and solid-line represent muon MDM varying with $ML_a$, for
$MLs=600{\rm GeV}$, $MLs=800{\rm GeV}$ and $MLs=1000{\rm GeV}$ respectively.}\label{oSLMLF}
\end{figure}

In order to embody the flavor mixing effects, we suppose $MLa\neq 0$. With $ALa=0$ and $MLs=600(800,1000)$ GeV,
   we plot the results with the varying $MLa$. Fig.\ref{oSLMLF} represents the relation between one loop scalar lepton-neutralino
   contribution and $MLs, MLa$.

The solid line is the result with $MLs=1000$ GeV, and $MLa$ varies from 0 to 450GeV.  $MLa$ affects the
 solid line weakly, but one can still find the result is increasing function of $MLa$. With $MLs=1000$GeV, the one loop scalar
 lepton-neutralino contribution to muon MDM is around $4.5\times10^{-10}$. The dashed line representing result with $MLa=800$ GeV,
 and the level influenced by $MLa$ is a little stronger than that of the solid line. The dashed line implies the result is
 about $8.0\times10^{-10}$. The dotted line is obtained with $MLs=600$ GeV, and it is strongly affected by $MLa$.
 $MLa=0$, the dotted line is about $14\times10^{-10}$. When $MLa>300$ GeV, the dotted line increases quickly. With $MLa=400$ GeV,
 the dotted line can reach $20\times10^{-10}$ and even larger. These results imply the flavor mixing can enhance
 the contributions. The flavor mixing enhance extent can be approximately expressed by the ratio between off-diagonal elements and
 diagonal elements for $(m_{\tilde{\nu}^c}^2)_{IJ},(m_L^2)_{IJ},(m_R^2)_{IJ}$, whose concrete form is $\frac{MLa^2}{MLs^2}$.

\begin{figure}[h]
\setlength{\unitlength}{1mm}
\centering
\includegraphics[width=4.0in]{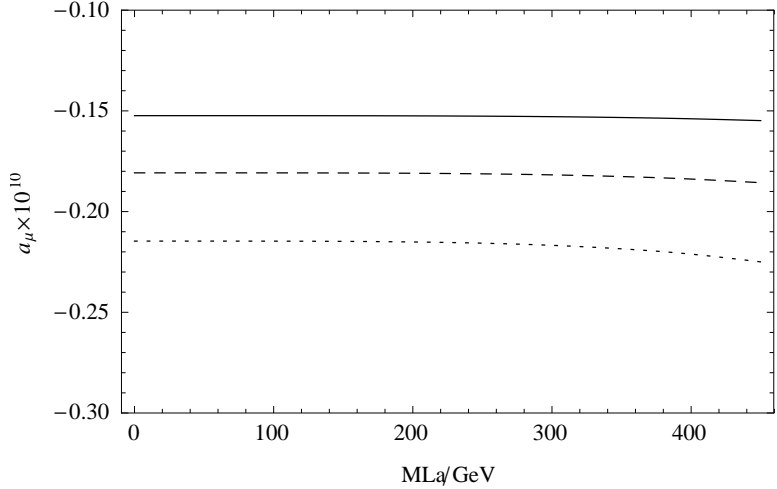}
\caption[]{The one loop scalar neutrino-chargino contributions to muon MDM,
 the dotted-line, dashed-line and solid-line represent muon MDM varying with $ML_a$, for
$MLs=600{\rm GeV}$, $MLs=800{\rm GeV}$ and $MLs=1000{\rm GeV}$ respectively.}\label{oSnuMLF}
\end{figure}
In Fig.\ref{oSnuMLF}, the one loop scalar neutrino-chargino contributions to muon MDM are obtained.
The solid line corresponds to $MLs=1000$ GeV, and the result is about $-1.5\times10^{-11}$. The dashed line result with
$MLs=800$ GeV is about $-1.7\times10^{-10}$. The dotted line representing $MLs=600$ GeV result, and it is around
$-2.2\times10^{-11}$. These three lines vary weakly with $MLa$. However, we also can find that the
extent affected by $MLa$ for the three lines follow the same rule: dotted-line $>$dashed-line$>$solid-line.

We also calculate the contribution from the off-diagonal element $ALa$. The numerical results imply that the effect of $ALa$
is very small for both the one loop scalar lepton-neutralino contribution and the scalar neutrino-chargino contribution.
Therefore, one can neglect $ALa$ safely, and in the latter numerical study, we suppose $ALa=0$.

\subsection{the sum of one loop and the two loop results}

\begin{figure}[h]
\setlength{\unitlength}{1mm}
\centering
\includegraphics[width=4.0in]{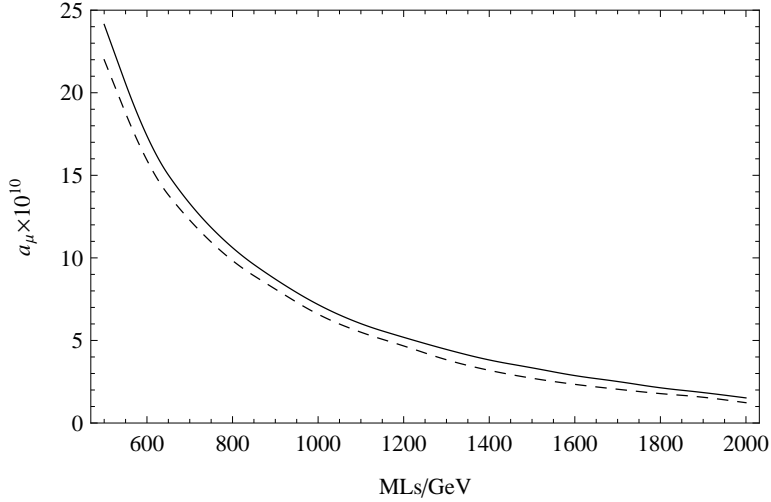}
\caption[]{As $AE_{45}=550{\rm GeV}$ and
$A_{cs}=-500{\rm GeV},MQ_2=1000{\rm GeV}$,
 the dashed-line and solid-line represent muon MDM varying with $ML_s$, for
$A_L=-500{\rm GeV}$ and $A_L=-800{\rm GeV}$ respectively.}\label{MLsNtu}
\end{figure}
Supposing, $AE_{45}=550{\rm GeV}$ and
$A_{cs}=-500{\rm GeV},MQ_2=1000{\rm GeV}$, in Fig.\ref{MLsNtu} we plot the muon MDM varying
with $ML_s$ for $A_L=-500{\rm GeV}$ and $A_L=-800{\rm GeV}$ respectively.
The solid line is on the dashed line. It implies when $A_L=-800{\rm GeV}$, the
numerical results are larger than the corresponding results with $AL=-500{\rm GeV}$.
When $ML_s<1000{\rm GeV}$, the numerical results turn large quickly with the decrescent $ML_s$.
In the $ML_s$ region $500 \sim 600{\rm GeV}$, the new physics contributions reach $20\times 10^{-10}$,
and the values can be even larger. The large value is able to remedy the deviation between
the SM prediction and experiment result for muon MDM well. With the enlarging $MLs$, the one loop
results decrease obviously. However, the two loop contributions having no relation with $MLs$ do
not change. That is to say the importance of the two loop contributions turns large when the
one loop contributions decrease with the enlarging $MLs$.

In Ref.\cite{one-loopsusy}, authors study some two loop diagrams in MSSM, where a
loop of charginos or neutralinos, the superpartners of gauge and Higgs, is
inserted into a two-Higgs-doublet one-loop diagram. Their numerical results can reach
$10\times10^{-10}$, which is large. They also study the two loop SUSY corrections
to muon MDM from the diagrams with a closed scalar fermion or fermion loop and/or
Higgs boson exchange. These contributions are in the region of $0.5\sigma\sim3\sigma$.
Our two loop results are at the order of $10^{-10}$. With the used parameters in BLMSSM, our studied
two loop results vary in the region of $0.5\sim 4.0\times10^{-10}$.

When the sub-scalar loop particles are Higgs(charged Higgs) and the virtual vectors are $\gamma,Z,W$,
these type two loop contributions are small $\sim 10^{-14}$ and even smaller.
The scalar neutrino loop and exotic scalar neutrino loop contributions are at the order of
$10^{-11}$. The contributions from the scalar leptons and scalar quarks
are of $10^{-11}\sim10^{-10}$ order. The exotic scalar quark contributions
is at the order of $10^{-12}\sim10^{-11}$, and it is smaller than the exotic scalar lepton
contributions $10^{-11}\sim10^{-10}$.

For the sub-Fermion loop, the corrections from the SM particles $(\tau,b,t)$ are small $10^{-15}\sim10^{-12}$, because of
the Fermion-Fermion-Higgs coupling. In this condition, to obtain considerable contributions the fermion should
be heavy to enhance the Fermion-Fermion-Higgs coupling. However, when the virtual particles are all very
heavy, their contributions will be suppressed. In the numerical results  the two loop neutralino and chargino contributions
are at the order of $10^{-11}\sim 10^{-10}$. While, the contributions from the exotic leptons, exotic quarks and exotic
neutrinos are at the order of $10^{-12}\sim10^{-11}$.

\begin{figure}[h]
\setlength{\unitlength}{1mm}
\centering
\includegraphics[width=4.0in]{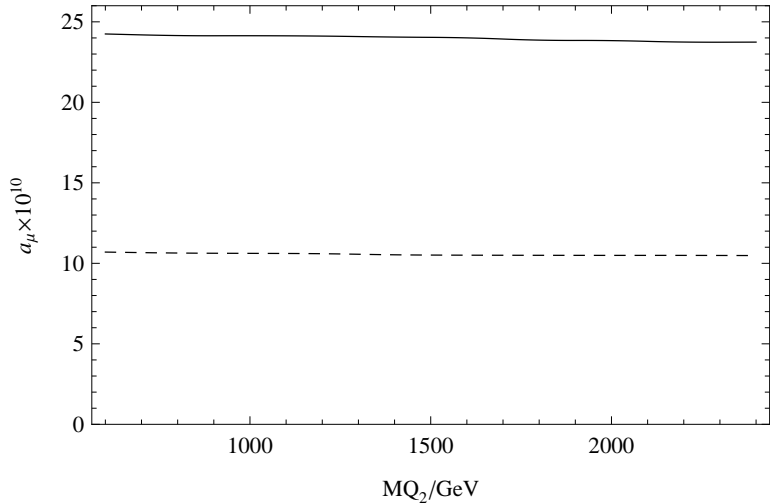}
\caption[]{As $AE_{45}=550{\rm GeV}$ and
$A_{cs}=-500{\rm GeV},AL=-800{\rm GeV}$,
 the solid-line and dashed-line represent muon MDM varying with $MQ_2$, for
$ML_s=500{\rm GeV}$ and $ML_s=800{\rm GeV}$ respectively.}\label{MQSNtu}
\end{figure}

The parameters $MQ_2$ relates with the square mass matrix of the 2nd
generation scalar quarks, and it's contribution to muon MDM appears at two loop level.
Therefore, it's effect should smaller than that of $ML_s$. Taking $AE_{45}=550{\rm GeV}$,
$A_{cs}=-500{\rm GeV}$, $A_L=-800{\rm GeV}$ and $ML_s=500(800){\rm GeV}$, the numerical results are obtained with
 in Fig.\ref{MQSNtu}, which shows the muon MDM varying with $MQ_2$ very mildly. The dashed line represents the result for $ML_s=800{\rm GeV}$, and is about $10.5\times10^{-10}$. On the other hand, the result shown as the solid line is around $24\times10^{-10}$.
The BLMSSM corrections decrease weakly with the enlarging $MQ_2$.

\begin{figure}[h]
\setlength{\unitlength}{1mm}
\centering
\includegraphics[width=4.0in]{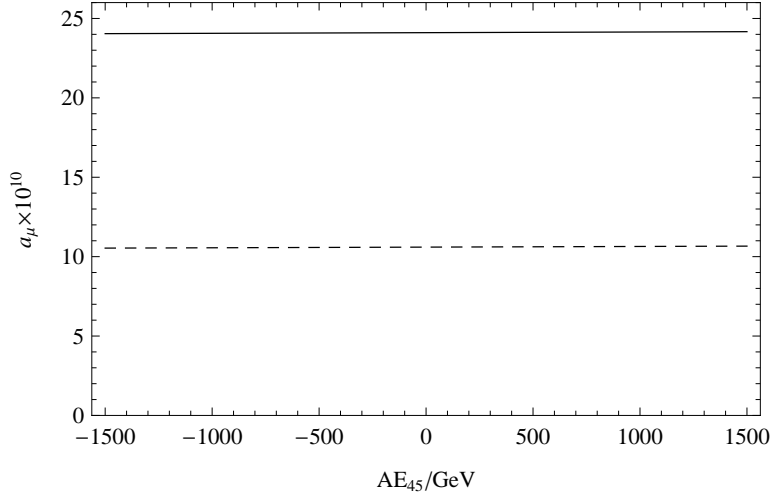}
\caption[]{As
$A_{L}=-800{\rm GeV}, A_{cs}=-500{\rm GeV},MQ_2=1000{\rm GeV}$,
 the  solid-line and dashed-line represent muon MDM varying with $AE_{45}$, for
$ML_{s}=500{\rm GeV}$ and $ML_{s}=800{\rm GeV}$ respectively.}\label{AE45tuN}
\end{figure}

The squared mass matrixes of the charged exotic scalar leptons
contain the parameters $AE_{45}$. With $A_{cs}=-500{\rm GeV},MQ_2=1000{\rm GeV},A_{L}=-800{\rm GeV}$,
 we plot the results versus $AE_{45}$ for $ML_2=500{\rm GeV}$
and $ML_2=800{\rm GeV}$ respectively. From Fig.\ref{AE45tuN}, one finds that the $AE_{45}$ affects the results slightly.
When $ML_2=800{\rm GeV}$, the corrections are about $10.5\times10^{-10}$. As while as, the corrections
reach $24\times10^{-10}$ with $ML_2=500{\rm GeV}$. The $AE_{45}$ effect to muon MDM is in the region $10^{-12}\sim10^{-11}$. However,
we still can see that the correction is the very slowly increasing function of $AE_{45}$.

\begin{figure}[h]
\setlength{\unitlength}{1mm}
\centering
\includegraphics[width=4.0in]{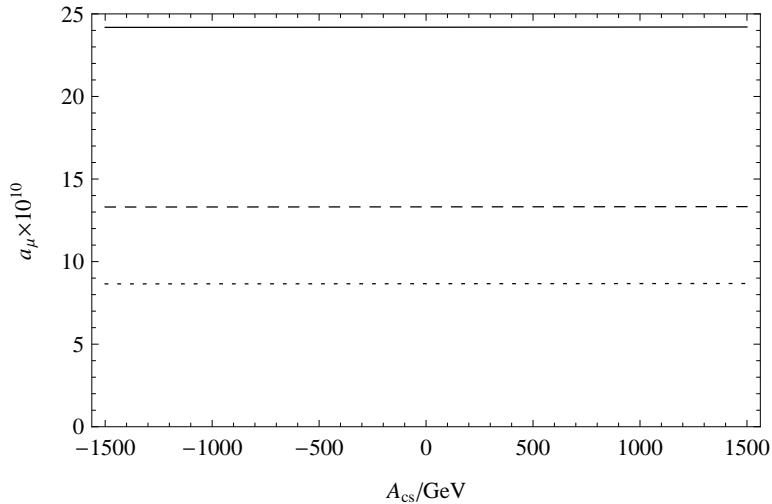}
\caption[]{As $AE_{45}=550{\rm GeV},A_L=-800{\rm GeV},MQ_2=1000{\rm GeV}$,
 the solid-line, dashed-line and dotted line represent muon MDM varying with $A_{cs}$, for
$ML_s=500(700,900){\rm GeV}$ respectively.}\label{ACStuN}
\end{figure}

In Fig.\ref{ACStuN}, we plot the results versus $A_{cs}$ for $ML_s=500(700,900){\rm GeV}$
with $AE_{45}=550{\rm GeV},AL=-800{\rm GeV}, MQ_2=1000{\rm GeV}$. The solid line is obtained
with $ML_s=500 {\rm GeV}$, and the result is about $24\times10^{-10}$. The dashed line representing
the correction with $ML_s=700 {\rm GeV}$ is around $13.5\times10^{-10}$. For $ML_s=900 {\rm GeV}$, the
correction is about $8.5\times10^{-10}$. The three lines all turn weakly with the varying $A_{cs}$.
From Figs.(\ref{MQSNtu},\ref{AE45tuN},\ref{ACStuN}), one can easily find that the parameters just
relating with the two loop contributions to muon MDM have small influence to the results, because
the one loop contribution is dominant.
\pagebreak[4]
\section{discussion and conclusion}

In the framework of the BLMSSM, the muon MDM is studied in this work. We calculate the one loop diagrams and the Barr-Zee type two loop diagrams. In the numerical analysis, we consider the experiment constraints such as:  the experiment data of the lightest CP-even Higgs and neutrino.
Our numerical results imply when the exotic and SUSY particles are not very heavy such as at TeV scale, the new physics
contribution is about $8.0\times10^{-10}$ and even larger. In the parameter space as we supposed, our
numerical results can reach $24\times10^{-10}$, as scalar leptons at $500{\rm GeV}$ scale, which can well remedy the deviation between the
experiment data and the SM theoretical prediction for muon MDM.

\appendix

  \section{the functions}
  The one-loop, two-loop functions and the form factors are collected here.
 {\footnotesize \begin{eqnarray}
&&\mathcal{C}(x,y,z)=-\frac{1}{16\pi^2}\Big(\frac{x\log (x)}{ (x-y)
   (x-z)}+\frac{y\log (y) }{(y-x) (y-z)}+\frac{z\log (z)}{ (x-z) (y-z)}\Big)\nonumber\\&&
\mathcal{C}_1(x,y,z)=\frac{1}{32 \pi ^2}\Big(\frac{x \log ^2(x)}{ (x-y)
   (x-z)}+\frac{y \log ^2(y)}{ (x-y)
   (z-y)}+\frac{z  \log ^2(z)}{
    (z-x) (z-y)}\Big)\nonumber\\&&
\mathcal{F}(v, f,t, s, w)=\frac{1}{512 \pi ^4}\Big(\frac{\mathcal{G}(f,s,w)}{(f-t)
   (f-v)}+\frac{\mathcal{G}(t,s,w)}{
   (f-t) (v-t)}+\frac{
   \mathcal{G}(v,s,w)}{ (v-f)
   (v-t)}\Big)
   \nonumber\\&&
\mathcal{G}(x,y,z)=-\Phi(x,y,z)-2 (x+y+z)+2 (x \log (x)+y
   \log (y)+z \log (z))\nonumber\\&&\hspace{2.2cm}-x \log ^2(x)-y \log
   ^2(y)-z \log ^2(z).
   \nonumber\\&&
   \mathcal{A}(x)=-\frac{x \log (x)}{16 \pi ^2},~~~\mathcal{F}_1(v, f,t, s)= \frac{\mathcal{G}(s,f,t)- \mathcal{G}(v,f,t)}
   {512 \pi ^4 (s-v)}.\\
  &&\mathcal{W}_{1}(v,f,t,s,w)=\frac{1}{24}\Big((4w-t)
   \frac{\partial^2}{\partial t\partial s}+(4s-t)
   \frac{\partial^2}{\partial t\partial w}- \frac{\partial}{\partial s} -
   \frac{\partial}{\partial w}\Big)\mathcal{F}_1(f,s,w,t)+\frac{1}{8}\frac{\partial}{\partial s} \mathcal{F}_1(f,s,w,v)\nonumber\\&&
   +
   \frac{1}{24}[(- t +4 w -4 s)\frac{\partial\mathcal{A}(s)}{\partial s}+(4 s
- t -4 w
   )\frac{\partial\mathcal{A}(w)}{\partial w}-\frac{7 t}{16 \pi ^2}]\frac{\partial}{\partial t}\mathcal{C}(v,f,t)
   + \frac{1}{24}[(2-3 s
   \frac{\partial}{\partial s})\frac{\partial\mathcal{A}(s)}{\partial s}-\frac{7}{16 \pi
   ^2}\nonumber\\&&-(3 w
   \frac{\partial}{\partial w}+1)\frac{\partial\mathcal{A}(w)}{\partial w}]\mathcal{C}(v,f,t)+(1-2t\frac{\partial}{\partial t})\frac{\mathcal{C}_1(v,f,t)}{384
   \pi ^2}
   +\frac{1}{24}\Big( [s (t+4 v+8 w-4s)-t v-t w-4 w^2]\nonumber\\&&\times
   \frac{\partial^2}{\partial t\partial w}+ (s (8 w-t-4s)-(4 w-t) (w-v))
   \frac{\partial^2}{\partial t\partial s}+4 t
   \frac{\partial}{\partial t}+7+ (s-v-7 w)
   \frac{\partial}{\partial w}+3 (s-w)(w
   \frac{\partial^2}{\partial w^2}\nonumber\\&&- s
   \frac{\partial^2}{\partial s^2})+ (3 t-4 s-v-2 w)
   \frac{\partial}{\partial s}\Big)\mathcal{F}(v,f,t,s,w)
   \\&&
    \mathcal{W}_{2}(v,f,t,s,w)=
   \frac{1}{72}[(1+2 s
   \frac{\partial}{\partial s})\frac{\partial\mathcal{A}(s)}{\partial s}+(1+2 w
   \frac{\partial}{\partial w})\frac{\partial\mathcal{A}(w)}{\partial w}]\mathcal{C}(v,f,t)
   +\frac{1}{48}\Big([\frac{2}{3}+ (t-
   5s - w  )\frac{\partial}{\partial s}]\nonumber\\&&\times\frac{\partial\mathcal{A}(s)}{\partial s}
+[\frac{2}{3}+ (t
- s -5 w )\frac{\partial}{\partial w}]\frac{\partial\mathcal{A}(w)}{\partial w}-\frac{1}{3\pi ^2}\Big)
   \frac{\partial\mathcal{C}(v,f,t)}{\partial t}
   +\frac{1}{72}\Big([(4 w-  t-4 s)\frac{\partial\mathcal{A}(s)}{\partial s}+(4 s
   -t-4 w)\nonumber\\&&\times\frac{\partial\mathcal{A}(w)}{\partial w}-\frac{t}{4 \pi
   ^2}]  (\frac{\partial^2}{\partial t^2}+\frac{\partial^2}{\partial v\partial t})\Big)\mathcal{C}(v,f,t)+\frac{1}{576 \pi
   ^2}(\frac{\partial}{\partial t}
   -t \frac{\partial^2}{\partial t^2}-t \frac{\partial^2}{\partial v\partial t}+\frac{\partial}{\partial v})\mathcal{C}_1(v,f,t)
+\frac{1}{144} \Big([2+( w\nonumber\\&&
-7 s  -v )\frac{\partial}{\partial s}]\frac{\partial\mathcal{A}(s)}{\partial s}+[2+( s
- v-7 w)\frac{\partial}{\partial w}]\frac{\partial\mathcal{A}(w)}{\partial w}-\frac{1}{2 \pi ^2}\Big)
   \frac{\partial\mathcal{C}(v,f,t)}{\partial v}
   +\frac{1}{144}\Big(5\frac{\partial}{\partial t} (\frac{\partial}{\partial s}+
   \frac{\partial}{\partial w})+2 (4 w\nonumber\\&&-t)
   \frac{\partial^3}{\partial t^2\partial s}+3(
   \frac{\partial^2}{\partial s^2}+
   \frac{\partial^2}{\partial w^2}+s
   \frac{\partial^3}{\partial s^3}+w
   \frac{\partial^3}{\partial w^3})+2 (4 s-t)
   \frac{\partial^3}{\partial t^2\partial w}+3 (t-3 s-w)
   \frac{\partial^3}{\partial t\partial s^2}+3 (t-s-3 w)\nonumber\\&&\times
   \frac{\partial^3}{\partial t\partial w^2}\Big)\mathcal{F}_1(f,w,s,t)
   +\frac{1}{144}\Big(3
   \frac{\partial^2}{\partial s\partial v}-
   \frac{\partial^2}{\partial s^2}+3
   \frac{\partial^2}{\partial w\partial w}-
   \frac{\partial^2}{\partial w^2}+ (3 s-v+w)
  \frac{\partial^3}{\partial s^2\partial v}+ (s-v+3 w)\nonumber\\&&\times
   \frac{\partial^3}{\partial v\partial w^2}\Big)\mathcal{F}_1(f,w,s,v)
   +\frac{1}{144}\Big(4
    \frac{\partial}{\partial t}-4
    \frac{\partial}{\partial s}-4
    \frac{\partial}{\partial w}-3 w (s-v+3 w)
    \frac{\partial^3}{\partial w^3}+ (3 v-2 s-t-7 w)
    \frac{\partial^2}{\partial w^2}\nonumber\\&&+ (3 v-7 s-t-2 w)
   \frac{\partial^2}{\partial s^2}-3 s (3 s-v+w)
    \frac{\partial^3}{\partial s^3}+2 [(8 w-t-4 s)
   s-(w-v) (4 w-t)] (\frac{\partial^3}{\partial t\partial s\partial v}+\frac{\partial^3}{\partial t^2\partial s})\nonumber\\&&+ (-5 t+5 v-28 w)
    \frac{\partial^2}{\partial t\partial w}+3 [(4 w-t-v+s) s-5 w^2+t
   v+(t-3 v) w]  \frac{\partial^3}{\partial t\partial w^2}+ (5
   v-28 s-5 t)  \frac{\partial^2}{\partial t\partial s}\nonumber\\&&+3 ((t-3 v+4 w-5 s)
   s+(w-t) (w-v))  \frac{\partial^3}{\partial t\partial s^2}+4
    \frac{\partial}{\partial v}+ (3 t-v-12 w)
    \frac{\partial^2}{\partial v\partial w}+ (3 t-12 s-v)
    \frac{\partial^2}{\partial v\partial s}\nonumber\\&&+ ((t+v+8 w-s) s-7 w^2-t
   v+3 t w-v w)  \frac{\partial^3}{\partial v\partial w^2}+ ((3 t-v+8 w-7 s) s-(w-t)
   (w-v))  \frac{\partial^3}{\partial v\partial s^2}\nonumber\\&&-4 t
    \frac{\partial^2}{\partial t^2}-4 t
    \frac{\partial^2}{\partial t\partial v}+2 (-4 s^2+(t+4 v+8 w) s-4 w^2-t
   v-t w)  (\frac{\partial^3}{\partial t^2\partial w}+ \frac{\partial^3}{\partial t\partial w\partial v})\Big)\mathcal{F}(v,f,t,s,w)
   \\
      &&\mathcal{W}_{3}(f,v,t,s,w,Q_{S_1},Q_{S_2})=
(\frac{\partial\mathcal{A}(s)}{\partial s}+\frac{1}{8\pi
   ^2})\frac{\mathcal{C}(f,v,t)}{24}+[( v-2 s
   +2 w )\frac{\partial\mathcal{A}(s)}{\partial s}+\frac{v}{8
   \pi ^2}-14 \mathcal{A}(s)+14  \mathcal{A}(w)]\nonumber\\&&\times
   \frac{\partial}{\partial v}
   \frac{\mathcal{C}(f,v,t)}{24}+\frac{1}{8}( \mathcal{A}(w)-
   \mathcal{A}(s))(2t   \frac{\partial^2}{\partial v\partial t}
   + v\frac{\partial^2}{\partial v^2}
   + t\frac{\partial^2}{\partial t^2}
)\mathcal{C}(f,v,t)+\frac{1}{384 \pi
   ^2}[(w-s)(6t \frac{\partial^2}{\partial t\partial v}+3(v+t)
    \frac{\partial^2}{\partial t^2})\nonumber\\&&+1
  +(v-16 s+16 w)
   \frac{\partial}{\partial v}]\mathcal{C}_1(f,t,v)
  +\frac{1}{24}[(v -6 s  +2w   )\frac{\partial}{\partial v}+1
]\frac{\partial}{\partial s}\mathcal{F}_1(t,s,w,v)
   +\frac{1}{24}\Big(1 +   (f+s-w)
   \frac{\partial}{\partial s}\nonumber\\&&+   (v-16 s+16 w)
   \frac{\partial}{\partial v}+
   (f (v-6 s+2 w)-(2 s-v-2 w) (s-w))
   \frac{\partial^2}{\partial s\partial v}+3(w-s)[2
   t   \frac{\partial^2}{\partial t\partial v}+
   v  \frac{\partial^2}{\partial v^2}+
  t    \frac{\partial^2}{\partial t^2}]\Big)\nonumber\\&&\times\mathcal{F}(f,v,t,s,w)
  +Q_{S_2}   \Big\{  \frac{1} {24}[(1+3 w   \frac{\partial}{\partial w})\frac{\partial\mathcal{A}(w)}{\partial w}+\frac{1}{8 \pi
   ^2}]\mathcal{C}(f,v,t)+\frac{1}{6}[(\frac{1}{4}
   t-   s + w  )\frac{\partial\mathcal{A}(w)}{\partial w}+\frac{t}{32\pi
   ^2}-\mathcal{A}(s)\nonumber\\&&+\mathcal{A}(w)]\frac{\partial}{\partial t}
   \mathcal{C}(f,v,t)+\frac{1}{384 \pi^2}((-8 s+t+8 w)
   \frac{\partial}{\partial t}+1   )\mathcal{C}_1(f,v,t)+[(\frac{1}{24} t-\frac{1}{6} s)
   \frac{\partial^2}{\partial t\partial w}
  -\frac{1}{8} w   \frac{\partial^2}{\partial w^2}-\frac{5}{24}
   \frac{\partial}{\partial w}]\nonumber\\&&\times\mathcal{F}_1(v,w,s,t)+\frac{1}{8}( w
     \frac{\partial}{\partial w}+2) \frac{\partial}{\partial w}\mathcal{F}_1(f,w,s,t)+\frac{1}{24}[1
  + (-5 f-s+6 v+7 w)    \frac{\partial}{\partial w}+   (-8 s+t+8 w)
    \frac{\partial}{\partial t}\nonumber\\&&+3   w (-f-s+v+w)    \frac{\partial^2}{\partial w^2}+
   (f (t-4 s)+(4 s-t-4 w) (s-w)) \frac{\partial^2}{\partial t\partial w}]\mathcal{F}(f,v,t,s,w)\Big\}
   +Q_{S_1} \Big\{s\leftrightarrow w\Big\}\\&&
\mathcal{W}_5(z,f,s,x,y)=\mathcal{W}_{11}(z,f,s,x,y)+(-\frac{1}{24}
   \frac{\partial}{\partial x}+\frac{1}{3}
   \frac{\partial}{\partial y})\mathcal{F}_1(f,x,y,s)-\frac{1}{8} \frac{\partial}{\partial y}
   \mathcal{F}_1(f,y,x,z)+\Big(\frac{1}{24} (-3
   s\nonumber\\&&+x-7 y+8 z)
   \frac{\partial}{\partial y}+\frac{1}{24}
   (-7 x+y-z)
   \frac{\partial}{\partial x} \Big)\mathcal{F}(z,f,s,x,y)
   \nonumber\\&&
 \mathcal{W}_6(z,f,s,x,y)=\mathcal{W}_{11}(z,f,s,y,x) +(\frac{5}{24}
   \frac{\partial}{\partial x}-\frac{5}{12}
   \frac{\partial}{\partial y})\mathcal{F}_1(f,x,y,s)+\frac{3}{8} \frac{\partial}{\partial y}
   \mathcal{F}_1(f,y,x,z)+\Big(\frac{1}{24} (9s\nonumber\\&&+x-7 y-10z)
   \frac{\partial}{\partial y}+\frac{1}{24}   (-7x+y+5z)
   \frac{\partial}{\partial x} \Big)\mathcal{F}(z,f,s,x,y)\\&&
   \mathcal{W}_7(z,f,s,x,y)=
   \Big(\frac{5}{36}
   \frac{\partial^2}{\partial x\partial s}-\frac{5}{72}
   \frac{\partial^2}{\partial y\partial s}+(\frac{y}{18}
   -\frac{s}{18})   \frac{\partial^3}{\partial x\partial s^2}+(\frac{1}{36} s+\frac{1}{18}x)
   \frac{\partial^3}{\partial y\partial s^2}+\frac{1}{24} (s-x)
   \frac{\partial^3}{\partial y^2\partial s}\nonumber\\&&+\frac{1}{24}
   \frac{\partial^2}{\partial y^2}+\frac{1}{24}
   y \frac{\partial^3}{\partial y^3}
   -\frac{1}{8} x \frac{\partial^3}{\partial s\partial x^2}
\Big)\mathcal{F}_1(f,x,y,s)+\frac{1}{72} \Big((x+6
   y-z) \frac{\partial^3}{\partial y^2\partial z}+6
   \frac{\partial^2}{\partial y\partial z}-3
   \frac{\partial^2}{\partial x\partial z}-
   \frac{\partial^2}{\partial y^2}\nonumber\\&&-3x \frac{\partial^3}{\partial x^2\partial z}
\Big)\mathcal{F}_1(f,x,y,z)+\frac{\mathcal{C}(z,f,s)}{12}
   (\frac{1}{3}+\frac{1}{2} y
   \frac{\partial}{\partial y}) \frac{\partial^2\mathcal{A}(y)}{\partial y^2}+\frac{1}{144} \Big(   (2-18 x
   \frac{\partial}{\partial x})\frac{\partial\mathcal{A}(x)}{\partial x}+[2 +(6 s-6 x\nonumber\\&&
   -12 y )\frac{\partial}{\partial y}]\frac{\partial\mathcal{A}(y)}{\partial y}-\frac{1}{\pi   ^2}\Big)
   \frac{\partial \mathcal{C}(z,f,s)}{\partial s}+\Big(\frac{\partial}{\partial z}-s
   \frac{\partial^2}{\partial s^2}
   +\frac{\partial}{\partial s}-s\frac{\partial^2}{\partial z\partial s}\Big)\frac{\mathcal{C}_1(z,f,s)}{576 \pi ^2}
   -\Big(\frac{s}{288\pi^2}+\frac{1}{18}
   (s+x\nonumber\\&&-y) \frac{\partial\mathcal{ A}(x)}{\partial x}-\frac{1}{36} (s+2 x-2 y)
   \frac{\partial\mathcal{ A}(y)}{\partial y}\Big)
\frac{\partial}{\partial s}(\frac{\partial}{\partial s}+\frac{\partial}{\partial z})\mathcal{C}(z,f,s)+\frac{1}{288} \Big(-4(1+3 x \frac{\partial}{\partial x}) \frac{\partial\mathcal{ A}(x)}{\partial x}
+[12+(4 x\nonumber\\&&-16 y -4 z)\frac{\partial}{\partial y}]
   \frac{\partial\mathcal{ A}(y)}{\partial y}-\frac{1}{\pi ^2}\Big)
   \frac{\partial\mathcal{C}(z,f,s)}{\partial z}+\Big(\frac{1}{24}
   \frac{\partial}{\partial y}+\frac{1}{36}
   \frac{\partial}{\partial s}+\frac{1}{72}   (-s-2 x-7 y+3 z)
   \frac{\partial^2}{\partial y^2}-\frac{1}{24}   y (x+y\nonumber\\&&-z)
   \frac{\partial^3}{\partial y^3}-\frac{7}{72}
   \frac{\partial}{\partial x}+\frac{1}{72}   (8 s-3 x-11 y-5 z)
   \frac{\partial^2}{\partial s\partial y}+\frac{1}{24}
   [x^2+(y-z) x-2 y^2+s (-x+y+z)]
   \frac{\partial^3}{\partial s\partial y^2}\nonumber\\&&+\frac{1}{72}
   (-13 s-17 x+3 y+10 z)
   \frac{\partial^2}{\partial s\partial x}+\frac{1}{36}
   [s (-x+y+z)-2 (x^2-(2 y+z) x+y^2)]
   \frac{\partial^3}{\partial s^2\partial y}+\frac{1}{18}
   (-x^2\nonumber\\&&-s x+2 y x-y^2+s y-s z+y z)
   \frac{\partial^3}{\partial s^2\partial x}+\frac{1}{72}
   (6 s-x-5 y-3 z)
   \frac{\partial^2}{\partial z\partial y}+\frac{1}{72}
   (s (x+6 y-z)-(x-y) (x\nonumber\\&&-4 y-z))
   \frac{\partial^3}{\partial z\partial y^2}+\frac{1}{72}
   (-3 s-7 x+y+2 z)   \frac{\partial^2}{\partial z\partial x}+\frac{1}{36}
   [s (-x+y+z)-2(x^2-(2 y+z) x+y^2)]
   \frac{\partial^3}{\partial z\partial y\partial s}\nonumber\\&&-\frac{1}{36}   s
   \frac{\partial^2}{\partial s\partial z}-\frac{1}{36}   s
   \frac{\partial^2}{\partial s^2}+\frac{1}{36}
   \frac{\partial}{\partial z}+\frac{1}{18}
   (-x^2-s x+2 y x-y^2+s y-s z+y z)
   \frac{\partial^3}{\partial z\partial x\partial s}-\frac{1}{12} x^2\frac{\partial^3}{\partial x^3}\nonumber\\&&-\frac{1}{8} x(x-y+z)
   \frac{\partial^3}{\partial x^2\partial s}
-\frac{1}{24} x(s+x-y)   \frac{\partial^3}{\partial z\partial x^2}
\Big)\mathcal{F}(z,f,s,x,y)   \nonumber\\&&
 \mathcal{W}_8(z,f,s,x,y)=\frac{1}{8}
   \Big[-\frac{\partial}{\partial x}
   \mathcal{F}_1(f,x,y,s)-
   \frac{\partial}{\partial y}
   \mathcal{F}_1(f,x,y,z)+(
   \frac{\partial \mathcal{A}(x)}{\partial x}-\frac{\partial \mathcal{A}(y)}{\partial y})
   \mathcal{C}(z,f,s)+\Big((x-s\nonumber\\&&-y)
   \frac{\partial}{\partial y}+(x
   -y-z) \frac{\partial}{\partial x}\Big)\mathcal{F}(z,f,s,x,y)\Big]\nonumber\\&&
  \mathcal{W}_9(z,f,s,x,y)= \frac{1}{8}\Big[-\frac{\partial}{\partial x}
   \mathcal{F}_1(f,x,y,s)-
   \frac{\partial}{\partial y}
   \mathcal{F}_1(f,x,y,z)+(
   \frac{-\partial \mathcal{A}(x)}{\partial x}+\frac{\partial \mathcal{A}(y)}{\partial y})
   \mathcal{C}(z,f,s)+\Big((y-s\nonumber\\&&-x)\frac{\partial}{\partial y}+(-x+y-z)
   \frac{\partial}{\partial x}\Big)\mathcal{F}(z,f,s,x,y)\Big].\\&&
\mathcal{W}_{10}(z,f,s,x,y)= \frac{1}{4}\Big[(\frac{\partial \mathcal{A}(x)}{\partial x}
   -\frac{\partial \mathcal{A}(y)}{\partial y})
  (\frac{\partial }{\partial s}+\frac{\partial }{\partial z})\mathcal{C}(z,f,s)- \frac{\partial^2 \mathcal{A}(x)}{\partial x^2}
  \mathcal{C}(z,f,s)-\Big(   \frac{\partial^2}{\partial x\partial s}+
   \frac{\partial^2}{\partial y\partial s}\nonumber\\&&+
   \frac{\partial^2}{\partial x^2}\Big)\mathcal{F}_1(f,x,y,s)
+\Big(\frac{\partial}{\partial x}-
   \frac{\partial}{\partial y}-2 y
   \frac{\partial^2}{\partial y^2}+(x+y-z)
   \frac{\partial^2}{\partial x^2}+(x-y-z)[
   \frac{\partial^2}{\partial y\partial s}+
   \frac{\partial^2}{\partial x\partial s}+
   \frac{\partial^2}{\partial y\partial z}\nonumber\\&&+
   \frac{\partial^2}{\partial z\partial x}]\Big)\mathcal{F}(z,f,s,x,y)\Big] \nonumber\\&&
    \mathcal{W}_{11}(z,f,s,x,y)=\Big(\frac{1}{6} (y-s)
   \frac{\partial^2}{\partial x\partial s}+\frac{1}{12} (s+2x)
   \frac{\partial^2}{\partial y\partial s}-\frac{1}{8} x
   \frac{\partial^2}{\partial x^2}+\frac{1}{8} y
   \frac{\partial^2}{\partial y^2}\Big)\mathcal{F}_1(f,x,y,s)-(s\frac{
   \partial}{\partial s}+1)\nonumber\\&&\times\frac{\mathcal{C}_1(z,f,s)}{
   192 \pi ^2}-\frac{\mathcal{C}(z,f,s)}{24}\Big((1+3
   x\frac{\partial}{\partial x} )\frac{\partial\mathcal{A}(x)}{\partial x}+(1+3
   y\frac{\partial}{\partial y} )\frac{\partial\mathcal{A}(y)}{\partial y}+\frac{1}{4\pi
   ^2}\Big)+\Big(\frac{y-s-x}{6}
   \frac{\partial\mathcal{A}(x)}{\partial x}-\frac{s}{96 \pi ^2}\nonumber\\&&+\frac{1}{12} (s+2 x-2 y)
   \frac{\partial\mathcal{A}(y)}{\partial y}\Big)
   \frac{\partial\mathcal{C}(z,f,s)}{\partial s}+\Big(\frac{1}{8} y   (x-y+z)
   \frac{\partial^2}{\partial y^2}-\frac{1}{6}   ((s-2 y+x) x+(y-s) (y-z))
  \frac{\partial^2}{\partial s\partial x}\nonumber\\&&-\frac{1}{12}-\frac{1}{8} x
   (x-y+z)  \frac{\partial^2}{\partial x^2} -\frac{1}{12}
   s   \frac{\partial}{\partial s}+\frac{1}{12}   ((4 y-s+2 z-2 x) x-2 y^2+s y+s z)
  \frac{\partial^2}{\partial s\partial y}\Big)\mathcal{F}(z,f,s,x,y)\\&&
\mathcal{W}_{12}(z,f,s,x,y,Q_{F_1},Q_{F_2})=(\frac{11}{32
   \pi ^2}-\frac{\partial\mathcal{A}(x)}{\partial x})\frac{\mathcal{C}(z,f,s)}{12}+[\frac{3 s}{128 \pi
   ^2}-\frac{\mathcal{A}(x)}{2}+\frac{\mathcal{A}(y)}{2}]
    \frac{\partial}{\partial s}\mathcal{C}(z,f,s)+\frac{s}{8}[\frac{s}{32\pi
   ^2} \nonumber\\&&- \mathcal{A}(x)+ \mathcal{A}(y) ]
   \frac{\partial^2\mathcal{C}(z,f,s)}{\partial s^2}+\frac{1}{12}[(y-
  z- x )\frac{\partial\mathcal{A}(x)}{\partial x}+\frac{11 z}{32 \pi ^2}-7
 \mathcal{ A}(x)+7 \mathcal{A}(y)]
   \frac{\partial\mathcal{C}(z,f,s)}{\partial z}
   +\frac{1}{8}[\frac{ z}{32 \pi ^2}- \mathcal{A}(x)\nonumber\\&&+ \mathcal{A}(y)](2s
   \frac{\partial^2}{\partial z\partial s}
   +  z \frac{\partial^2}{\partial z^2})\mathcal{C}(z,f,s)+\frac{1}{384\pi^2}
   \Big(3(y-x)[ z \frac{\partial^2}{\partial z^2}+2s
   \frac{\partial^2}{\partial z\partial s}+4
   \frac{\partial}{\partial s}+s \frac{\partial^2}{\partial s^2}]
   -2-2(8 x-8 y+z)\nonumber\\&&\times\frac{\partial}{\partial z}\Big)\mathcal{C}_1(z,f,s)
   +\frac{1}{12} [(-3 x+y-z)
  \frac{\partial^2}{\partial z\partial x}- \frac{\partial}{\partial x}]\mathcal{F}_1(f,x,y,z)
   +\Big((-s-x+y)(\frac{1}{12}
   \frac{\partial}{\partial x}+\frac{1}{2}
   \frac{\partial}{\partial s}
  +\frac{1}{8} s \frac{\partial^2}{\partial s^2})\nonumber\\&&+\frac{1}{12} (8 y-8
   x-7 z) \frac{\partial}{\partial z}-\frac{1}{12}
   [(3 s-2 y+z-x) x+(y-s) (y-z)]   \frac{\partial^2}{\partial z\partial x}-\frac{1}{8}(x-y+z)(2s
    \frac{\partial^2}{\partial z\partial s}+ z
    \frac{\partial^2}{\partial z^2})\nonumber\\&&-\frac{1}{3}
\Big)\mathcal{F}(z,f,s,x,y)
   +Q_{F_2}   \Big[\Big(-\frac{1}{3}\frac{\partial}{\partial y}+\frac{1}{12} (-s-2 x)
   \frac{\partial^2}{\partial y\partial s}-\frac{1}{8} y
   \frac{\partial^2}{\partial y^2}\Big)\mathcal{F}_1(f,y,x,s)+\frac{1}{4}
   \frac{\partial}{\partial y}\mathcal{F}_1(f,y,x,z)\nonumber\\&&+\mathcal{C}(z,f,s)
   [(\frac{1}{8} y
    \frac{\partial}{\partial y}-\frac{1}{12})\frac{\partial\mathcal{A}(y)}{\partial y}+\frac{1}{192 \pi ^2}]
    +\frac{1}{12}[(2 y- s-2 x )\frac{\partial\mathcal{A}(y)}{\partial y} +\frac{s}{16\pi
   ^2}-2\mathcal{A}(x)+2\mathcal{A}(y)]\frac{\partial\mathcal{ C}(z,f,s)}{\partial y}
\nonumber\\&&-\frac{\mathcal{C}_1(z,f,s)}{192 \pi ^2}[(s+4 x-4 y)
  \frac{\partial}{\partial s}+1]
   +\Big(\frac{1}{12} (3 s+x+2 y-4 z)
   \frac{\partial}{\partial y}+\frac{1}{8} y (y-x-z)
   \frac{\partial^2}{\partial y^2}+\frac{1}{12} (4 y-s-4 x)
  \frac{\partial}{\partial s}\nonumber\\&&+\frac{1}{12} [(s-4 y-2 z +2
   x) x+2 y^2-s y-s z]   \frac{\partial^2}{\partial s\partial y}-\frac{1}{12} \Big)\mathcal{F}(z,f,s,x,y)\Big]
   +Q_{F_1} \Big[   [(\frac{1}{6}+\frac{1}{8} x)\frac{\partial \mathcal{A}(x)}{\partial x}+\frac{1}{192 \pi ^2}]\nonumber\\&&\times\mathcal{C}(z,f,s)
   +\frac{1}{6}[( s+ x
   - y )\frac{\partial \mathcal{A}(x)}{\partial x}+\frac{s}{32 \pi
   ^2}+\mathcal{A}(x)-\mathcal{A}(y)]
   \frac{\partial}{\partial s}\mathcal{C}(z,f,s)+\frac{\mathcal{C}_1(z,f,s)}{96 \pi
   ^2}[(s+2 x-2 y)\frac{\partial}{\partial s}+1]\nonumber\\&&
   +[\frac{1}{6} \frac{\partial}{\partial x}+\frac{1}{6} (s-y)
    \frac{\partial^2}{\partial x\partial s}+\frac{1}{8} x
    \frac{\partial^2}{\partial x^2}]\mathcal{F}_1(f,x,y,s)+\Big(\frac{1}{12} (5 x-2 y+2 z)
   \frac{\partial}{\partial x}+\frac{ x }{8} (x-y+z)
    \frac{\partial^2}{\partial x^2}+\frac{1}{6} (s+2 x\nonumber\\&&-2 y)
    \frac{\partial}{\partial s}+\frac{1}{6} [x^2+(s-2
   y) x+(y-s) (y-z)] \frac{\partial^2}{\partial s\partial x}+\frac{1}{6} \Big)\mathcal{F}(z,f,s,x,y)\Big]\\
 && \mathcal{W}_{13}(z,f,s,x,y,Q_{F_1},Q_{F_2}) =W_{12}(z,f,s,x,y,Q_{F_1},Q_{F_2})+Q_{F_1} \Big(-\frac{1}{4} \frac{\partial}{\partial x} \mathcal{F}_1(f,x,y,s)-
  \frac{1}{4}   \frac{\partial \mathcal{A}(x)}{\partial x}
  \mathcal{ C}(z,f,s)\nonumber\\&&-\frac{\mathcal{C}_1(z,f,s)}{64 \pi
   ^2}+\frac{1}{4}[(y-x-z)  \frac{\partial}{\partial x}-1]\mathcal{F}(z,f,s,x,y)\Big)+Q_{F_2} \Big(-\frac{1}{4}
  \frac{\partial}{\partial y} \mathcal{F}_1(f,y,x,z)+
  \frac{1}{4} \frac{\partial \mathcal{A}(y)}{\partial y}
  \mathcal{C}(z,f,s)\nonumber\\&&+\frac{\mathcal{C}_1(z,f,s)}{64\pi^2}
+\frac{1}{4}[(-s-x+y)\frac{\partial}{\partial y}+1]\mathcal{F}(z,f,s,x,y)\Big)\\&&
 \mathcal{ W}_{14}(z,f,s,x,y,Q_{F_1},Q_{F_2}) =(\frac{\partial\mathcal{A}(x)}{\partial x}-\frac{7}{34 \pi
   ^2}) \frac{\mathcal{C}(z,f,s)}{6}+(\frac{\mathcal{A}(y)}{2}-\frac{3 s}{128 \pi
   ^2}-\frac{\mathcal{A}(x)}{2})
   \frac{\partial\mathcal{C}(z,f,s)}{\partial s}+\frac{s}{8}[\mathcal{A}(y)\nonumber\\&&- \mathcal{A}(x) -\frac{s}{32 \pi
   ^2}]\frac{\partial^2}{\partial s^2}\mathcal{C}(z,f,s)+\frac{1}{12}[(2
   z- x  + y )\frac{\partial\mathcal{A}(x)}{\partial x}-\frac{7 z}{32 \pi ^2}-7
   \mathcal{A}(x)+7 \mathcal{A}(y)]
   \frac{\partial}{\partial z}\mathcal{C}(z,f,s)+\frac{1}{8}(\mathcal{A}(y)\nonumber\\&&-\frac{ z}{32 \pi
   ^2}- \mathcal{A}(x))[2s
  \frac{\partial^2}{\partial z\partial s}+z
   \frac{\partial^2}{\partial z^2}]+\frac{1}{384 \pi ^2}\Big((y-x)[3 z
  \frac{\partial^2}{\partial z^2}+12
   \frac{\partial}{\partial s}+3s
   \frac{\partial^2}{\partial s^2}+6s
   \frac{\partial^2}{\partial z\partial s}]+4+4(-4 x\nonumber\\&&+4 y+z)
   \frac{\partial}{\partial z}\Big)\mathcal{C}_1(z,f,s)
   +[\frac{1}{12} (-3 x+y+2
   z)  \frac{\partial^2}{\partial z\partial x}+\frac{1}{6} \frac{\partial}{\partial x}]\mathcal{F}_1(f,x,y,z)+\Big(\frac{1}{6} (s+x-y)
  \frac{\partial}{\partial x}+(s-x\nonumber\\&&+y)[\frac{1}{2}
   \frac{\partial}{\partial s}+\frac{1}{8} s
    \frac{\partial^2}{\partial s^2}]+(-x+y+z)[\frac{2}{3}
   \frac{\partial}{\partial z}+\frac{1}{4} s
   \frac{\partial^2}{\partial z\partial s}+\frac{1}{8} z
   \frac{\partial^2}{\partial z^2}]+\frac{1}{12} [(2 z-3
   s+2 y-x) x-(y-s) (y\nonumber\\&&+2 z)]
   \frac{\partial^2}{\partial x\partial z}+\frac{5}{12}
\Big) \mathcal{F}(z,f,s,x,y)
   +Q_{F_2}   \Big(   [(\frac{1}{6}+\frac{1}{8} y
    \frac{\partial}{\partial y}) \frac{\partial\mathcal{A}(y)}{\partial y}+\frac{1}{192 \pi ^2}]\mathcal{C}(z,f,s)
    +\frac{1}{6}[( s- x  + y ) \frac{\partial\mathcal{A}(y)}{\partial y}\nonumber\\&&+\frac{s}{32 \pi
   ^2}-\mathcal{A}(x)+\mathcal{A}(y)]
  \frac{\partial}{\partial s}\mathcal{C}(z,f,s)+\frac{1}{96 \pi ^2}[(s-2 x+2 y)
    \frac{\partial}{\partial s}+1]\mathcal{C}_1(z,f,s)+[\frac{1}{6} (s-x)
    \frac{\partial^2}{\partial y\partial s}+\frac{ y}{8}
    \frac{\partial^2}{\partial y^2}\nonumber\\&&+\frac{5}{12}  \frac{\partial}{\partial y}]\mathcal{F}_1(f,y,x,s)-\frac{1}{4}
    \frac{\partial}{\partial y}\mathcal{F}_1(f,y,x,z)+[\frac{1}{12} (5 y-3 s-2 x+5 z)
    \frac{\partial}{\partial y}+\frac{1}{8} y (y-x+z)
    \frac{\partial^2}{\partial y^2}+\frac{1}{6}+\frac{1}{6} (s\nonumber\\&&-2 x+2 y)
    \frac{\partial}{\partial s}+\frac{1}{6}[y^2-(s+2
   y+z-x) x+s y+s z]\frac{\partial^2}{\partial y\partial s}]\mathcal{F}(z,f,s,x,y)\Big)
   +Q_{F_1}   \Big( [(\frac{x}{8}
   \frac{\partial}{\partial x}-\frac{1}{12})\frac{\partial\mathcal{A}(x)}{\partial x}\nonumber\\&&+\frac{1}{192 \pi ^2}]\mathcal{C}(z,f,s)
+\frac{1}{12}[(2 x- s
   -2y )\frac{\partial\mathcal{A}(x)}{\partial x}+\frac{s}{16 \pi
   ^2}+2\mathcal{A}(x)-2\mathcal{A}(y)]
   \frac{\partial}{\partial s}\mathcal{C}(z,f,s)-((s-4 x+4 y)
   \frac{\partial}{\partial s}\nonumber\\&&+1)\frac{\mathcal{C}_1(z,f,s)}{192 \pi
   ^2}+[\frac{1}{12} (-s-2 y)
  \frac{\partial^2}{\partial x\partial s}-\frac{1}{8} x
   \frac{\partial^2}{\partial x^2}-\frac{1}{12}
   \frac{\partial}{\partial x}]\mathcal{F}_1(f,x,y,s)+[\frac{1}{12} (2 x+y-z)
   \frac{\partial}{\partial x}+\frac{1}{8} x (x-y\nonumber\\&&-z)
   \frac{\partial^2}{\partial x^2}+\frac{1}{12} (-s+4 x-4 y)
   \frac{\partial}{\partial s}+\frac{1}{12} [2
   x^2-(s+4 y) x+(s+2 y) (y-z)]
   \frac{\partial^2}{\partial x\partial s}-\frac{1}{12}]\mathcal{F}(z,f,s,x,y)\Big)\\&&
    \mathcal{W}_{15}(z,f,s,x,y,Q_{F_1},Q_{F_2})= W_{14}(z,f,s,x,y,Q_{F_1},Q_{F_2})+ Q_{F_1}\Big(\frac{1}{4}
   \frac{\partial \mathcal{A}(x)}{\partial x}
  \mathcal{ C}(z,f,s)-\frac{1}{4}\frac{\partial}{\partial x} \mathcal{F}_1(f,x,y,s)\nonumber\\&&+\frac{\mathcal{C}_1(z,f,s)}{64 \pi
   ^2}+\frac{1}{4}[(x-y-z)
   \frac{\partial}{\partial x}+1]\mathcal{F}(z,f,s,x,y)\Big)+Q_{F_2} \Big(-\frac{1}{4}
   \frac{\partial}{\partial y}\mathcal{ F}_1(f,y,x,z)
   -\frac{1}{4}\frac{\partial \mathcal{A}(y)}{\partial y}
   \mathcal{C}(z,f,s)\nonumber\\&&-\frac{\mathcal{C}_1(z,f,s)}{64 \pi
   ^2}+\frac{1}{4}[(-s+x-y) \frac{\partial}{\partial y}-1]\mathcal{F}(z,f,s,x,y)\Big)
   \end{eqnarray}}
\end{document}